\documentclass[epjc3]{svjour3}
\journalname{Eur. Phys. J. C}
                     
\usepackage{graphicx}
\usepackage{bm}
\usepackage{epsf}
\usepackage{rotating}
\usepackage{epsfig,graphics,rotate,color}
\usepackage{wrapfig}
\usepackage{array,hhline,dcolumn}%
\usepackage[normalem]{ulem}
\usepackage{color}
\usepackage{hyperref}
\usepackage{graphicx}
\usepackage{placeins}
\usepackage{multirow}
\usepackage{makecell}
\usepackage[labelformat=simple]{subcaption}

\usepackage{lineno}

\begin{document}

\title{Separation of track-- and shower--like energy deposits in ProtoDUNE-SP using a convolutional neural network}

%

\author{The DUNE Collaboration\\\\
       A.~Abed Abud\thanksref{Liverpool,CERN}
       \and B.~Abi\thanksref{Oxford}
       \and R.~Acciarri\thanksref{Fermi}
       \and M.~A.~Acero\thanksref{Atlantico}
       \and M.~R.~Adames\thanksref{Tecnologica }
       \and G.~Adamov\thanksref{Georgian}
       \and M.~Adamowski\thanksref{Fermi}
       \and D.~Adams\thanksref{Brookhaven}
       \and M.~Adinolfi\thanksref{Bristol}
       \and A.~Aduszkiewicz\thanksref{Houston}
% ...
}

\institute{
% ...
\pagebreak[0] Universidad del Atl\'antico, Barranquilla, Atl\'antico, Colombia\label{Atlantico}
% ...
\and\pagebreak[0] University of Bristol, Bristol BS8 1TL, United Kingdom\label{Bristol}
        \and\pagebreak[0] Brookhaven National Laboratory, Upton, NY 11973, USA\label{Brookhaven}
% ...
\and\pagebreak[0] CERN, The European Organization for Nuclear Research, 1211 Meyrin, Switzerland\label{CERN}
% ...
\and\pagebreak[0] Fermi National Accelerator Laboratory, Batavia, IL 60510, USA\label{Fermi}
% ...
\and\pagebreak[0] Georgian Technical University, Tbilisi, Georgia\label{Georgian}
% ...
\and\pagebreak[0] University of Houston, Houston, TX 77204, USA\label{Houston}
% ...
\and\pagebreak[0] University of Liverpool, L69 7ZE, Liverpool, United Kingdom\label{Liverpool}
% ...
\and\pagebreak[0] University of Oxford, Oxford, OX1 3RH, United Kingdom\label{Oxford}
% ...
\and\pagebreak[0] Universidade Tecnol\'ogica Federal do Paran\'a, Curitiba, Brazil\label{Tecnologica }
% ...
}


\maketitle

\begin{abstract}
Liquid argon time projection chamber detector technology provides high spatial and calorimetric resolutions on the charged particles traversing liquid argon. As a result, the technology has been used in a number of recent neutrino experiments, and is the technology of choice for the Deep Underground Neutrino Experiment (DUNE). In order to perform high precision measurements of neutrinos in the detector, final state particles need to be effectively identified, and their energy accurately reconstructed. This article proposes an algorithm based on a convolutional neural network to perform the classification of energy deposits and reconstructed particles as track-like or arising from electromagnetic cascades. Results from testing the algorithm on experimental data from ProtoDUNE-SP, a prototype of the DUNE far detector, are presented. The network identifies track-- and shower--like particles, as well as Michel electrons, with high efficiency. The performance of the algorithm is consistent between experimental data and simulation. 
\end{abstract}


\section{Introduction}
The ProtoDUNE single phase detector (ProtoDUNE-SP)~\cite{Abi:2020mwi,DUNE:2021hwx} is a prototype liquid argon time projection chamber (LArTPC) for the Deep Underground Neutrino Experiment (DUNE) far detector~\cite{Abi:2020wmh,dunetdr_volume4}. ProtoDUNE-SP is known as a single phase detector as it is operated entirely within liquid phase argon. The detector readout mechanism consists of six Anode Plane Assemblies (APAs), each containing three wire readout planes at angles of $\pm36^\circ$ and $0^\circ$ to the vertical, where the readout planes are denoted U, V and W, respectively. The U and V views are the induction views, meaning that charge is induced on the wires by drifting electrons, and the W-view wires collect the drifting electrons. The wires in each readout plane are spaced with approximately 5\,mm pitch and are read out at a rate of 2\,MHz. A full description of the detector is given in Ref.~\cite{DUNE:2021hwx}. ProtoDUNE-SP collected data from a positively-charged-particle beam at CERN~\cite{PhysRevAccelBeams.20.111001,PhysRevAccelBeams.22.061003} in autumn 2018, including charged pions, charged kaons, protons, muons and positrons recorded with momenta in the range from 0.3 to 7.0\,GeV/c. Additionally, since ProtoDUNE-SP is located on the Earth's surface, it is subject to a large flux of cosmic ray muons.

\begin{figure*}[htp!]
    \centering
    \includegraphics[width=0.95\textwidth]{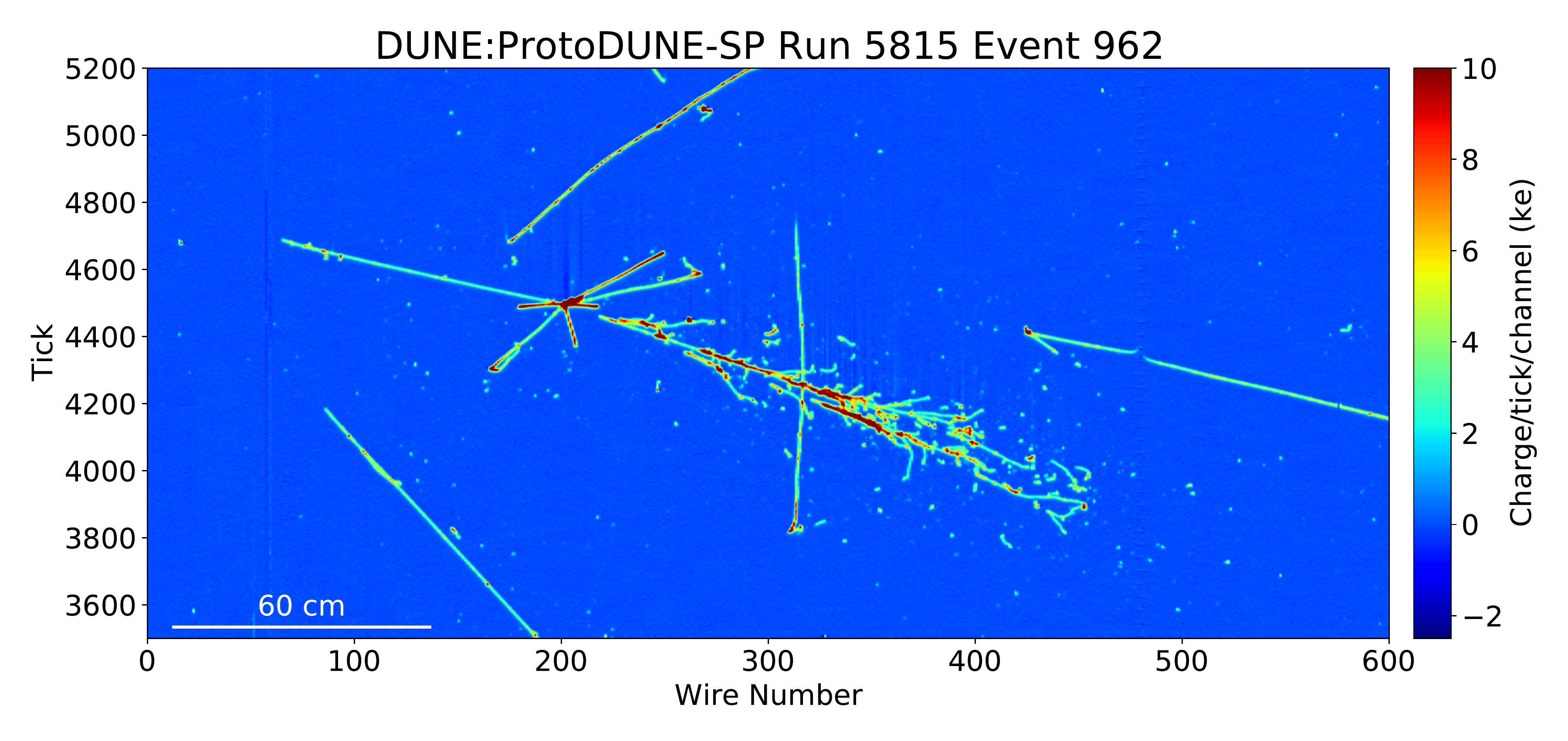}
    \caption{A 7\,GeV/$c$ beam $\pi^+$ interaction in the collection view (W-view) in ProtoDUNE-SP data. The x axis shows the wire number. The y axis shows the time tick in the unit of 0.5 $\mu$s. The colour scale represents the charge deposition.}
    \label{fig:r5815_e962}
\end{figure*}

The particle interactions can be easily and naturally visualised as three two-dimensional images (one for each readout view) in the wire number and time parameter space. Each pixel in the image represents the measured charge from a reconstructed energy deposition, called a \emph{hit}, on a given wire at a given time. A major challenge in the automated reconstruction of particle interactions in LArTPCs is identifying whether energy deposits originate from track-like (linear, such as protons, charged kaons, charged pions, and muons) or shower-like (locally dense, such as electrons and photons) structures. An example of a 7\,GeV/$c$ charged pion interaction is given in Fig.~\ref{fig:r5815_e962}, where the $\pi^+$ enters the detector and interacts (just after wire 200 and at time tick 4500) producing a number of track-- and shower--like particles. In order to classify the interaction type of the $\pi^+$, for example as charge-exchange or inelastic scattering, the particles emitted from the interaction vertex must be identified. 
The classification of reconstructed particles as track-like or shower-like will also be important in DUNE for the correct classification of neutrino interactions in the far detector. The identification of Michel electrons helps to distinguish between $\mu^{+}$ and $\mu^{-}$. It can also be used to identify stopping charged pions whose energy can be fully reconstructed.

In this article, we propose and demonstrate the use of a convolutional neural network (CNN) to classify hits as either belonging to track--like or shower--like structures~\cite{aidanthesis}. Furthermore, a Michel electron score is given to each hit to help identify Michel electrons and positrons, produced in the decay of muons and antimuons, respectively~\cite{Michel:1949qe}\footnote{Michel electron will be used to refer to both Michel electrons and Michel positrons.}. These hit-level classifications can be used alongside pattern recognition based reconstruction algorithms such as Pandora~\cite{pandorasdk,pandora} to refine the track or shower classification of reconstructed particles. The performance of the Pandora reconstruction on ProtoDUNE-SP simulated and experimental data is described in detail in Ref.~\cite{pandoraProtoDUNE}. Convolutional neural networks have been successfully used in neutrino physics for event classification~\cite{novacvn,microboonecvn,dunecvn}, particle identification~\cite{microbooneparticle,novaparticle} and reconstruction~\cite{Abbasi:2021ryj}. The fine-grain images obtained from LArTPC detectors makes CNNs a natural choice for such tasks. This algorithm is novel in that it aims to classify the hits based on a small local neighbourhood as opposed to a semantic segmentation approach that uses a much larger image containing a large part (or all) of the detector. The algorithm was designed in this way to minimise the memory usage and computational processing time, allowing it to run quickly on standard computing node CPUs where there is no access to powerful GPUs.

\section{The convolutional neural network}
Convolutional neural networks extract features from images by applying a series of filters that are learned during the training process~\cite{Jackel2008,Szegedy2015}. The number of filters and the number of convolutional layers varies for each specific use case; they are determined by the class of problem the network is trying to solve, and the computer hardware available for training and evaluating the network. In this case, a GPU was available for the training of the network, but the inference is performed on computing cluster CPUs (where GPUs are typically not available) as a part of the ProtoDUNE-SP reconstruction chain. As a result, only simple architectures containing few convolutional layers were considered, constrained by the desired evaluation time on the CPUs. For inference tasks within the ProtoDUNE-SP event reconstruction workflow, a C++ interface was added to the LArSoft framework~\cite{Church:2013hea}. Recent attempts to introduce GPU acceleration into the workflow mentioned above show promising reductions in processing time~\cite{GPUasaservice}.  

\begin{figure*}[htp!]
    \centering
    \includegraphics[width=0.95\textwidth]{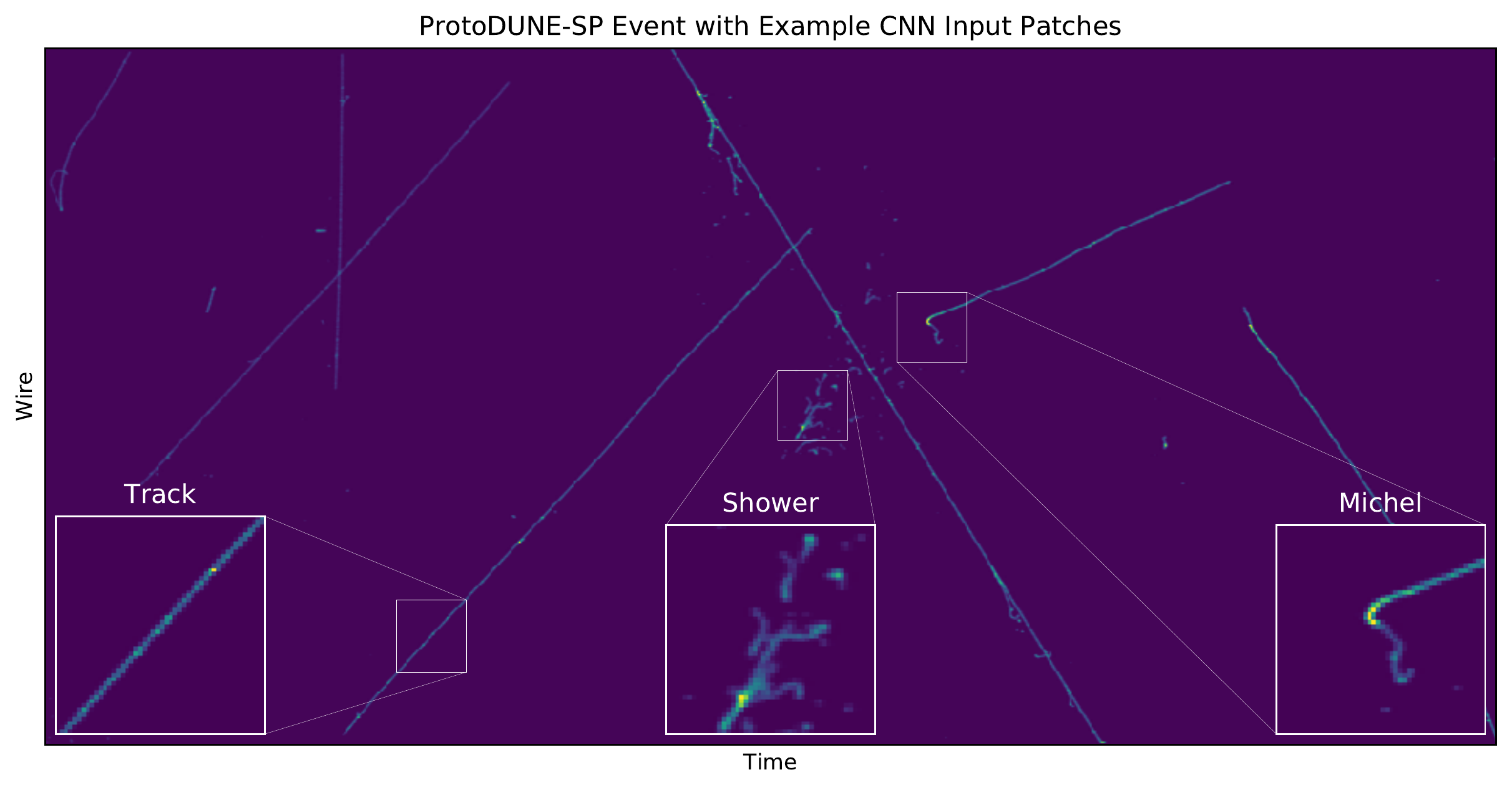}
    \caption{Examples of CNN input patches from a simulated ProtoDUNE-SP event. The inputs to the CNN are small $48\times 48$ pixel images created from patches of the full detector readout. Three examples are shown, each labelled with their appropriate class. The patch of the detector readout from which each patch was generated is emphasised.}
    \label{fig:patches}
\end{figure*}

The input to the network is a small region of the entire detector image known as a \emph{patch}. For each reconstructed hit object, the wire number $w$ and peak time $t$ are extracted, and a $48\times 48$ pixel image, centred on $\left(w,t\right)$, is created and the value of each pixel corresponds to the detected charge on a given wire at a given time. The wire dimension of the image corresponds to 48 wires with one wire per pixel. The time data are downsampled by averaging over six time samples, such that the spatial dimensions of the pixels match the 5\,mm wire pitch in both directions. Therefore, each image represents around $24\times 24$\,cm$^2$ of wire data. Figure~\ref{fig:patches} shows the hits from one APA in a simulated ProtoDUNE-SP event and the three zoomed regions give example $48\times 48$ pixel patches in the track, shower and Michel categories. Detector effects such as the ones introduced by space charge~\cite{Abi:2020mwi,Palestini:2021mlc} are included in the simulation. The images from the three wire planes are independently evaluated. This paper only reports on results from the collection plane, which has the highest signal-to-noise ratio. 

\begin{figure}[htp!]
    \centering
    \includegraphics[width=0.45\textwidth]{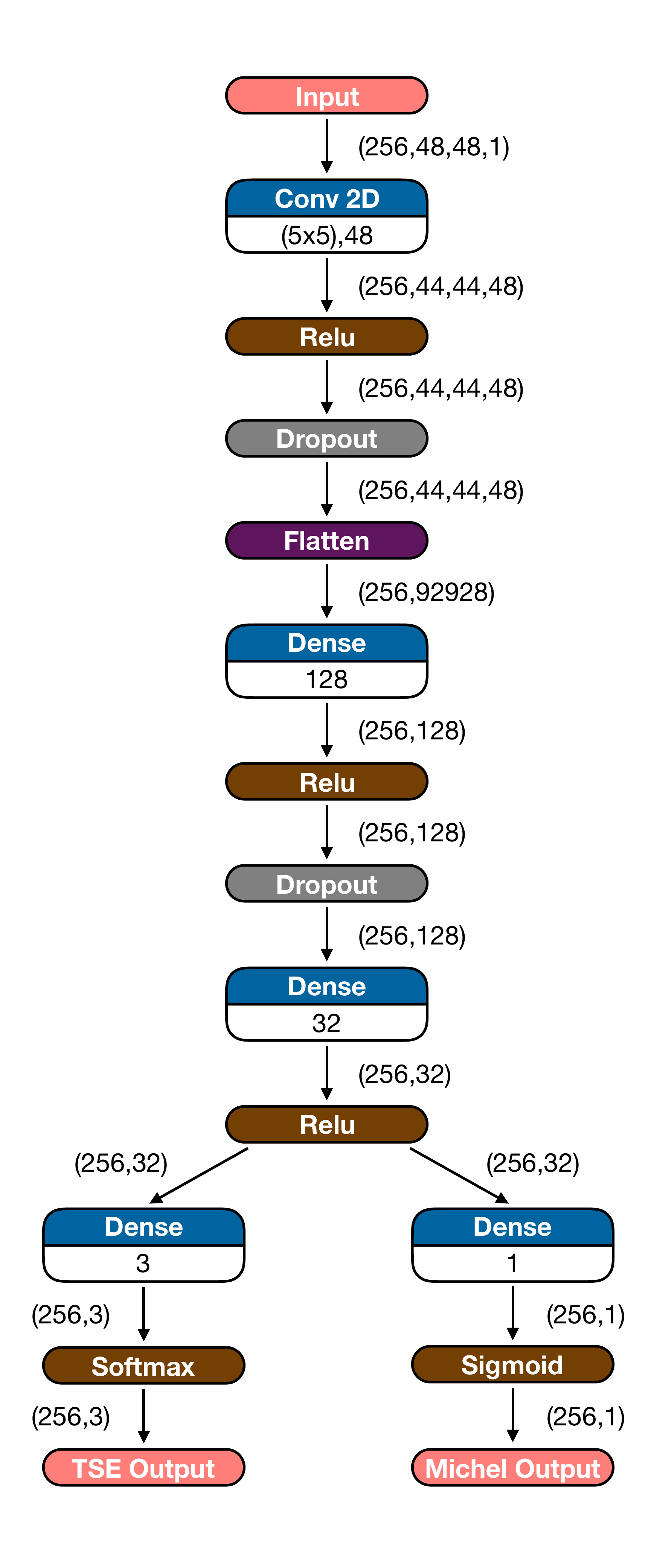}
    \caption{The CNN architecture. In this case, the CNN processes 256 images in parallel. Each image is a $48\times 48$ pixel patch of the calibrated detector readout. A single convolutional layer, with 48 filters of size $5\times 5$, is used to extract features from the images. These are processed by two dense layers containing 128 and 32 neurons respectively, before being split into two branches which provide the track-shower-empty (TSE) and Michel outputs. The dimensions of the data after each operation are given next to the black arrows.} 
    \label{fig:cnn_arch}
\end{figure}

The architecture for this hit-tagging CNN is shown in Fig.~\ref{fig:cnn_arch}. A single convolutional layer containing 48 $5\times 5$ pixel filters is used to extract feature maps from the input image, which are then flattened and passed to two dense layers that use them to classify the images. Two dropout\footnote{Dropout randomly disables a given fraction of neurons for each training example.} layers are used for regularisation~\cite{dropout}. The output of the network is split into two branches. The first branch returns the scores for track, shower, or empty (TSE) classification, which can be interpreted as probabilities as they are constrained to sum to one by a softmax~\cite{nwankpa2018activation} activation function. The second returns the probability for a Michel electron classification, with a sigmoid~\cite{nwankpa2018activation} activation function. The output of the network is split in this way due to the overlap of the shower and Michel electron classes. The total loss function is a weighted sum of the two branches, with the weights derived from the relative size of the training samples in each branch.


\subsection{Training details}
For the purposes of training a true classification must be attached to each of the patches. In addition to track, shower and Michel electron patches, empty patches are also created where the central pixel contains no energy deposit. Approximately 30 million images were prepared in total using approximately 500 simulated events: $\sim$15 million in the track sample, $\sim$11 million in the shower sample, $\sim$3 million in the empty sample, and $\sim$1 million in the Michel electron sample.

\begin{figure*}[htp!]
    \centering
    \includegraphics[width=0.95\textwidth]{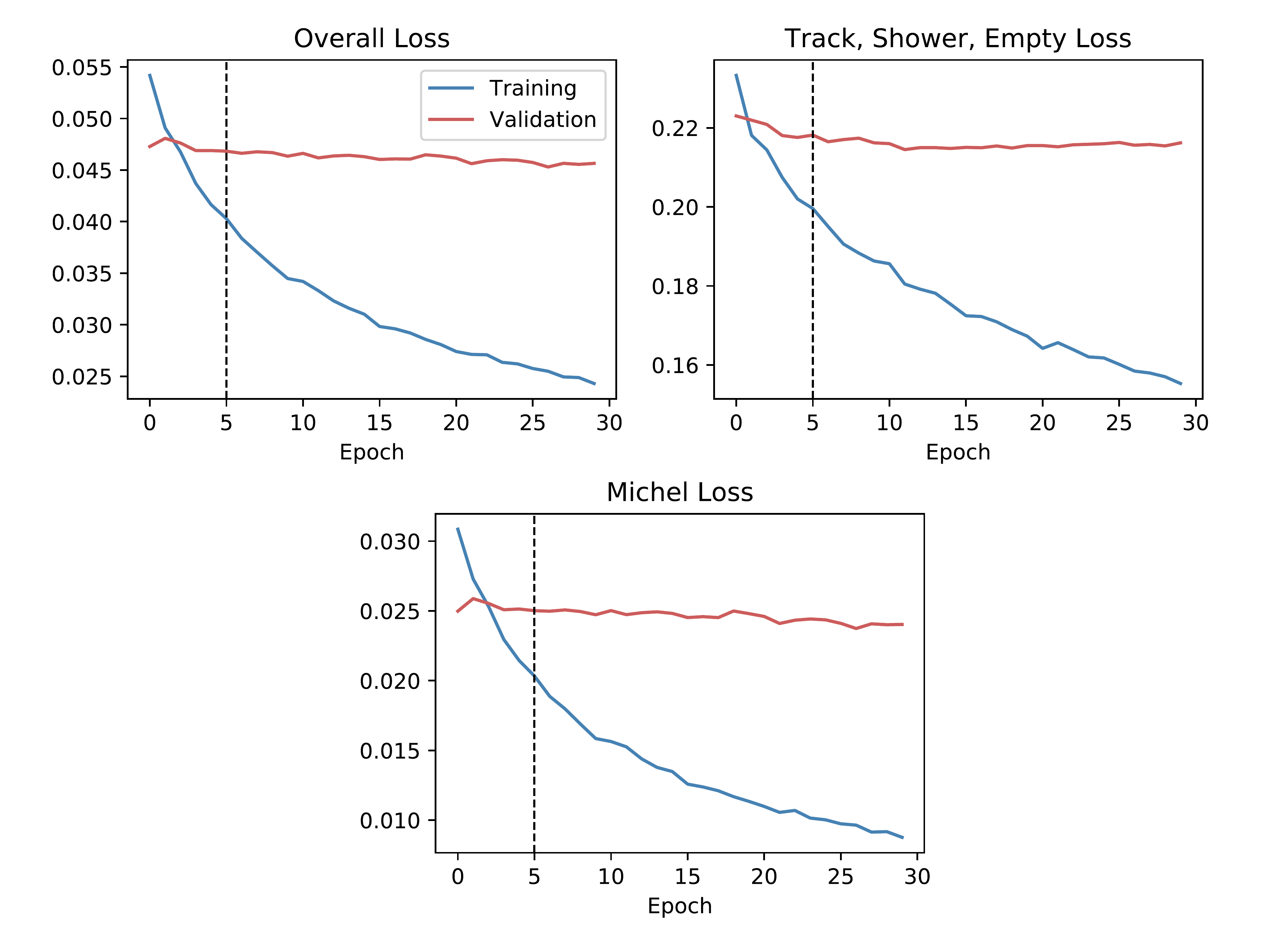}
    \caption{Evolution of the training and validation losses as a function of training epoch. The final weights of the network were taken from a checkpoint at the end of the fifth epoch, shown here as a vertical line. The overall loss; track, shower and empty loss; and Michel loss are shown in the top left, top right, and bottom left respectively. In calculating the overall loss, the track, shower and empty loss is weighted by 0.1 to be consistent with the smaller size of the Michel sample.}
    \label{fig:losses}
\end{figure*}

The CNN was trained with TensorFlow~\cite{Abadi-et-al-2016-tensorflow} through its Keras~\cite{chollet2015keras} interface, and performance metrics, such as the losses, purity and efficiency, were monitored throughout training using TensorBoard~\cite{tensorboard}. Before training, the data set was split into training, test, and validation sets in the ratio 80:10:10. The performance metrics were monitored throughout training with the training and validation sets, and again after training with the test set. Figure~\ref{fig:losses} shows the evolution of the training and validation losses throughout the training. Due to the large number of training images and relative simplicity of the task, the losses fall sharply within the first epoch, which is not visible in the plots. 
The training and validation losses begin to diverge after the first few epochs suggesting there is some over-fitting, but the network generalises well when considering the similar performance of the algorithm on the test and validation sets.
To further ensure generalisation, an early stopping algorithm was used, which focused on the loss in the TSE branch~\cite{Prechelt2012}. The final weights for the network were taken from a checkpoint at the end of the fifth epoch\footnote{An epoch is defined as one iteration over the entire training sample.} since the validation loss in the TSE branch starts to plateau on the fifth epoch.

\subsection{Performance}

\begin{figure}[htp!]
    \centering
    \includegraphics[width=0.8\textwidth]{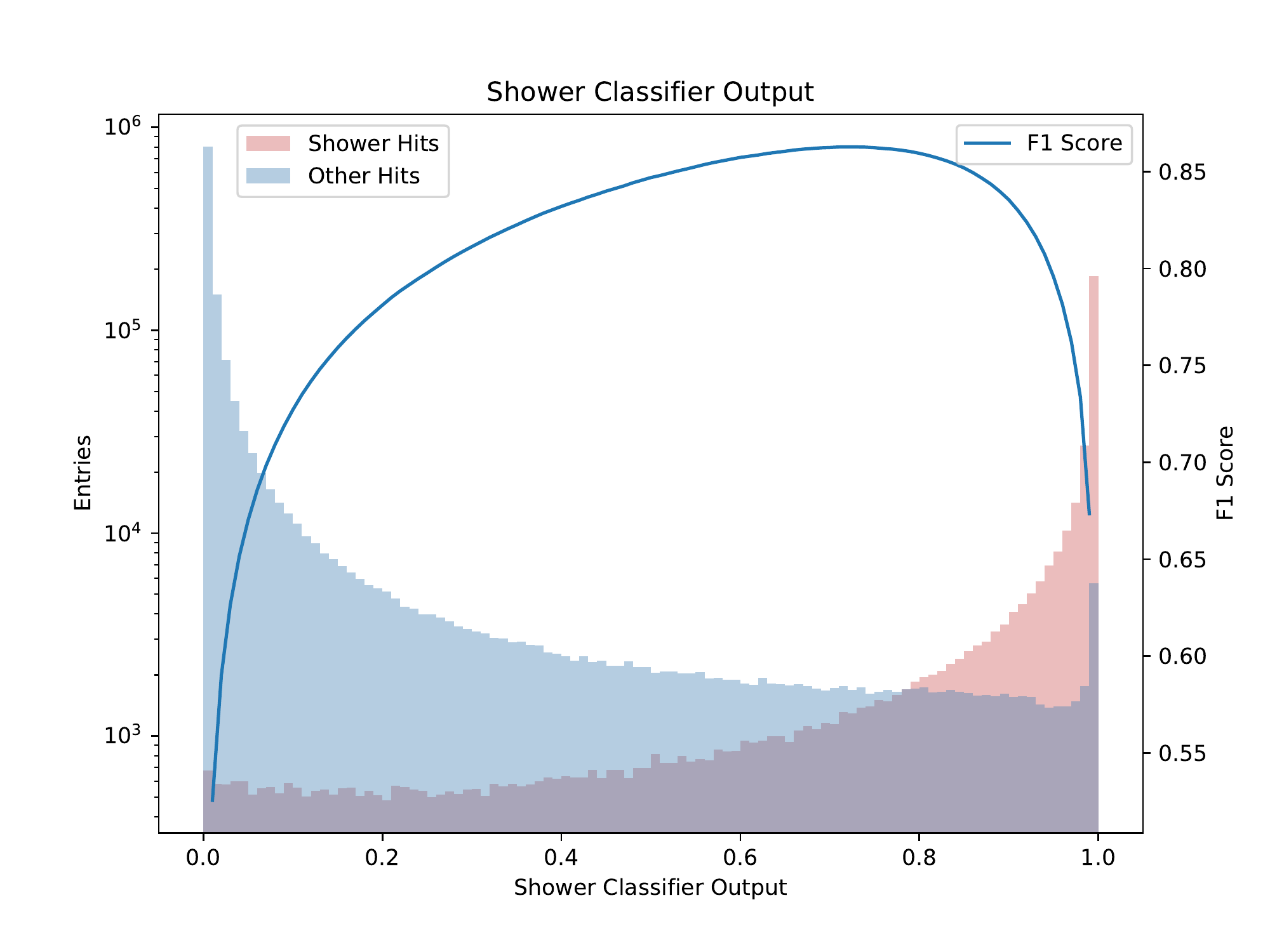}
    \caption{Shower classifier output distributions. The output of the shower classifier is shown for true shower hits in red and all other hits in blue. The blue line shows the F1 score as a function of classification threshold.}
    \label{fig:show_dist}
\end{figure}

The performance of the hit tagging was evaluated with reconstructed events from ProtoDUNE-SP simulation. A $48\times 48$ pixel image is created around each reconstructed hit, which is then classified by the network and the classification compared with the truth label. Note that by definition this method ensures that no processing is performed on empty images. Figure~\ref{fig:show_dist} shows the shower score distributions for true shower hits and all other hits, and a strong separation is seen between the distributions with a score close to one corresponding to a hit that is highly likely to come from a shower.
The small peak in the \textit{other hits} distribution close to one comes from delta-ray electrons overlapping with the cosmic-ray muon that produced them.  
The classification threshold can be set on a case by case basis, for the initial validation of the network on the ProtoDUNE-SP data it was optimised based on the F1 score, which is given by: 

\begin{equation}
    \frac{1}{F_{1}} = \frac{1}{2}\left( \frac{1}{\textrm{purity}} + \frac{1}{\textrm{efficiency}} \right),
\end{equation}
where the purity is defined as the fraction of correctly classified shower hits in the sample of all selected shower hits, and the efficiency as the fraction of all true shower hits that were selected as shower hits. 

\begin{figure}[htp!]
    \centering
    \includegraphics[width=0.8\textwidth]{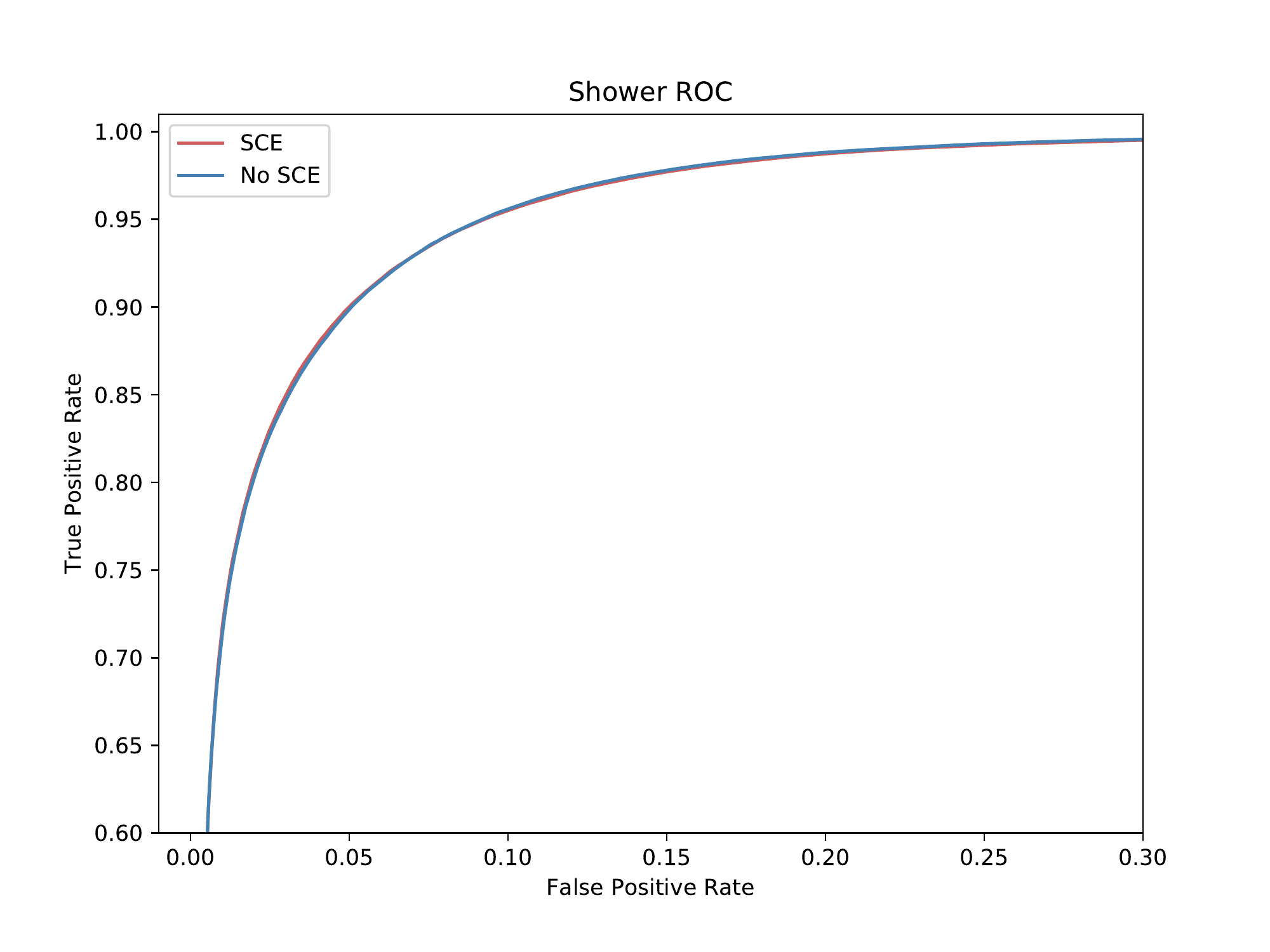}
    \caption{ROC curves for the shower classifier, showing the true positive rate against false the positive rate for varying classification threshold on the shower classifier output. The red (blue) line shows the ROC curve from ProtoDUNE-SP simulation with (without) SCE. The red curve is obscured by the blue due to close agreement.}
    \label{fig:show_roc}
\end{figure}

Figure~\ref{fig:show_roc} demonstrates the performance of the network in terms of the true positive and false positive rates. In this case, the true positive rate is the fraction of true shower hits that have been correctly classified as shower hits, and the false positive rate is the fraction of other hits incorrectly classified as shower hits. The receiver operating characteristic (ROC) curve is shown, which shows the true positive rate against the false positive rate as the selection threshold on the shower classifier output is varied. ROC curves are shown for simulation with the space charge effect (SCE, red) and without (blue). The close agreement between the curves suggests that the CNN results are robust against changes in the SCE model.

\begin{figure}[htp!]
    \centering
    \includegraphics[width=0.8\textwidth]{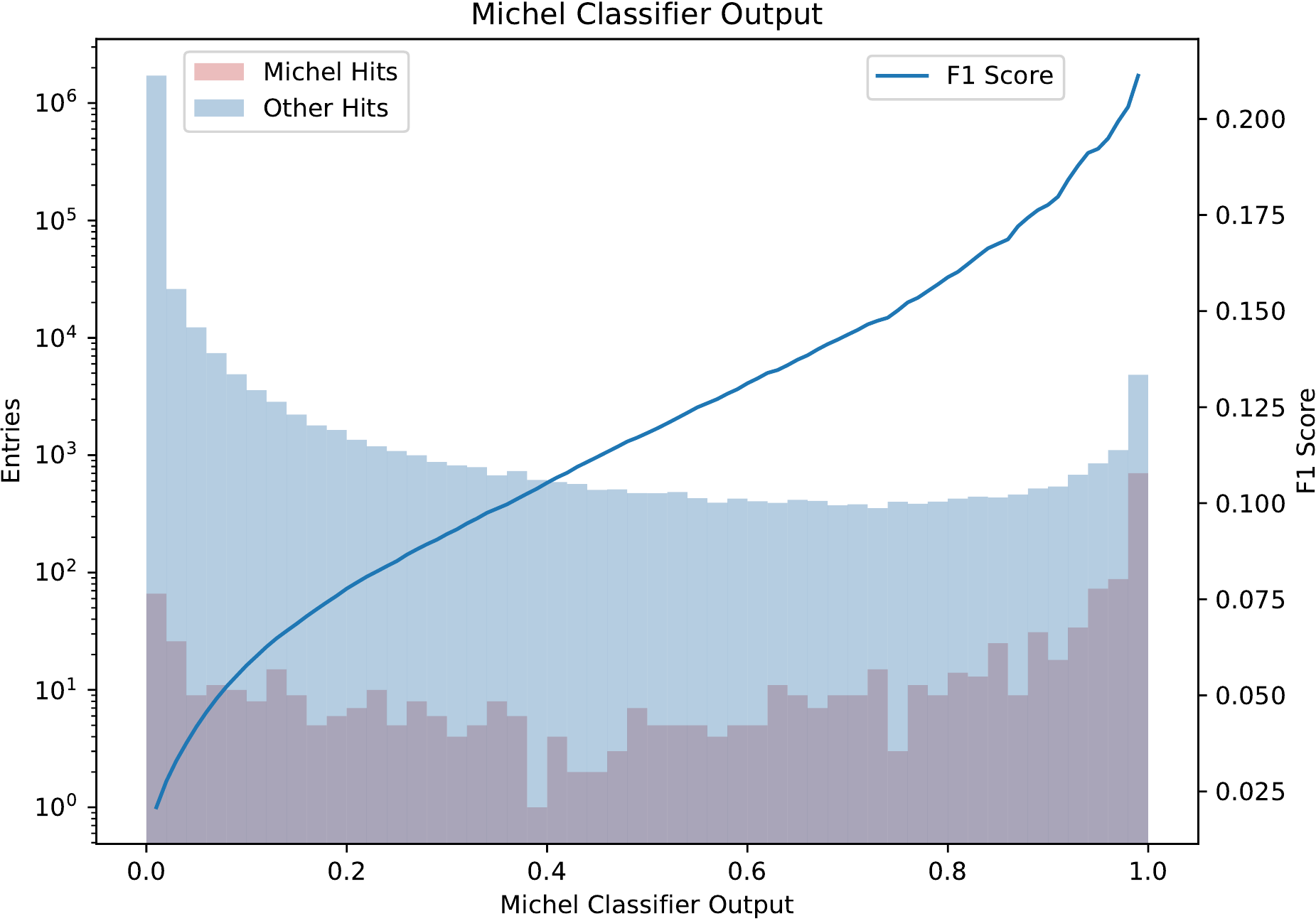}
    \caption{Michel electron classifier output distributions.}
    \label{fig:mich_dist}
\end{figure}

The score distributions from the Michel electron classifier are shown in Fig.~\ref{fig:mich_dist}, for true Michel electron hits and all other hits. 
The Michel electron classification is a significantly more challenging problem, partly due to the large variation in Michel electron interactions in the detector. Michel electrons can be seen as single short track-like objects, or more fragmented due to photon emission and subsequent Compton scattering to produce additional electrons. Furthermore, some delta-ray electrons and components of electromagnetic showers can produce signatures in the detector that are similar to Michel electrons. Therefore, while both distributions are strongly peaked at the expected values, with Michel electrons close to one and other hits close to zero, there are also sub-leading peaks of hits that are not correctly classified. Due to this, and the significantly smaller sample of Michel electron hits, the network is not able to achieve a good performance in terms of the F1 metric. However, when combined with simple clustering, a high purity sample of Michel electron events can be selected, as will be discussed in Sec.~\ref{sec:validation}.

\section{Results from experimental data and simulation}
\label{sec:validation}
It is important that the CNN is robust against potential differences between experimental data and simulation, and hence the performance has been compared between experimental data and simulation for several particle species. Hits are tagged in the three different readout views and reconstructed particles from Pandora are assigned a score between 0 and 1 that is the average of the shower classifier score from the CNN from all of the 2D hits in the collection view. Each hit is weighted by the hit charge when calculating the average shower score. A score close to one means that it is highly probable that the particle is shower--like, and a low score means the particle is very likely to be track--like. 

\begin{figure*}[htp!]
    \centering
    \includegraphics[width=0.95\textwidth]{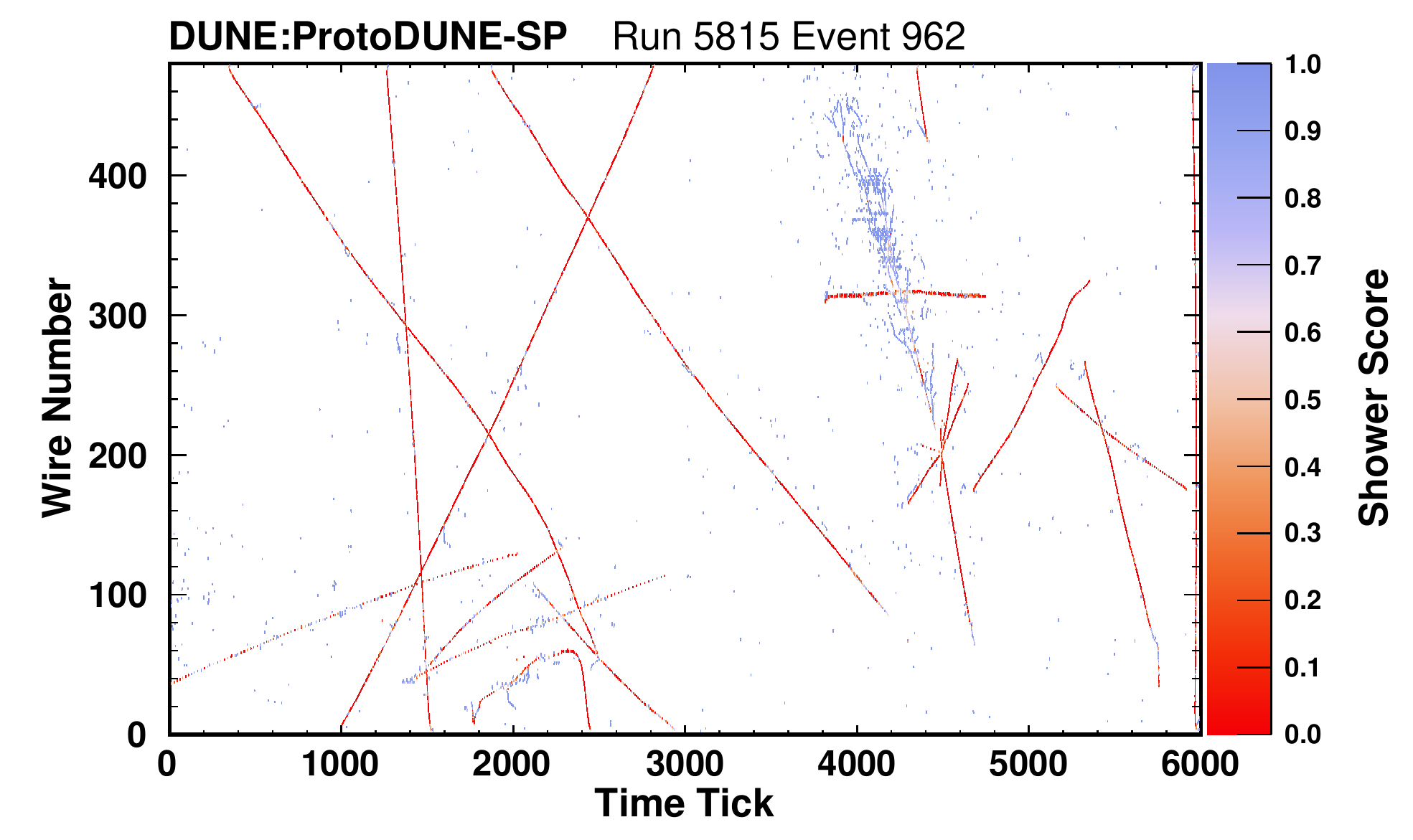}
    \caption{The CNN shower score of each hit in reconstructed particles for the same 7\,GeV/$c$ $\pi^+$ interaction shown in Fig.~\ref{fig:r5815_e962}. This diagram shows the location of reconstructed hits in wire--time coordinates, and the hits are coloured based on the CNN shower score. Red hits are track--like, and blue hits are shower--like. A number of cosmic muon tracks can be seen, along with tracks and showers produced by the pion interaction. The small shower--like patches along the muon tracks are delta-ray electrons.}
    \label{fig:event_r5387}
\end{figure*}

Data from ProtoDUNE-SP runs 5387 and 5809 taken in the H4-VLE test beam at CERN with 1\,GeV/$c$ beam momentum were used for the initial qualitative validation of the CNN performance on ProtoDUNE-SP data. These runs contain cosmic rays and particles from the charged particle beam. Run 5809 was taken with the inclusive beam trigger giving a dataset primarily consisting of beam positrons. Run 5387 was taken with a trigger that vetoed positrons, which resulted in a sample primarily consisting of beam $\pi^{+}$'s, $\mu^{+}$'s and protons. Figure~\ref{fig:event_r5387} shows an example of the CNN shower scores of reconstructed particles in a ProtoDUNE-SP event as a visual cross-check of the CNN performance. As expected, the cosmic-ray muon and pion tracks in the event have low shower scores, while the photon shower from the charged particle beam interaction is given a high score. In addition, delta ray electrons, which are emitted along the muon tracks, are associated with showers and therefore receive a high CNN shower score. The latest ProtoDUNE-SP Monte Carlo (MC) simulation sample was used to compare with experimental data. This is a new MC simulation with improved modelling of detector response, which is completely independent of the previous MC simulation that was used to train the CNN. 

The following sections report the performance of the CNN classification at the hit level and the particle level for cosmic rays and charged particles from the test beam. In order to classify the hits, a threshold of 0.72 was applied to the shower classifier output of the CNN, with hits exceeding the threshold being classified as shower hits. This threshold was selected by choosing the value with the largest F1 score in Fig.~\ref{fig:show_dist}. For particle-level classification, a different threshold of 0.81 is applied to the average shower score to classify particles, where the threshold was chosen to maximise the product of the selection efficiencies of all four types of charged beam particles.



\subsection{Cosmic-ray muons}

The CNN was first tested on a sample of cosmic-ray muons\footnote{Since ProtoDUNE-SP does not have a magnetic field this sample also contains antimuons.} in order to validate its performance before it was used to classify beam particles (see Sec.~\ref{sec:beam_results}).
A sample of cosmic-ray muons was selected from simulation and experimental data (run 5387), where cosmic-ray muon candidates were selected using the following criteria:
\begin{itemize}
    \item the particle was reconstructed by Pandora as a track
    \item the track was at least 1\,m in length
    \item the track started and ended at least 50\,cm from the front face of the detector (to veto beam particles)
    \item the track was directed at least $15^\circ$ away from the vertical (to veto tracks that only deposited energy on a small number of collection plane wires). 
\end{itemize}
All of the hits associated to the selected tracks were labelled as true cosmic-ray muon hits. The hits from any other particles associated with the cosmic-ray muon candidate, such as delta-ray and Michel electrons, were not included to avoid contaminating the hit selection.

\begin{figure}[htp!]
    \centering
    \includegraphics[width=0.48\textwidth]{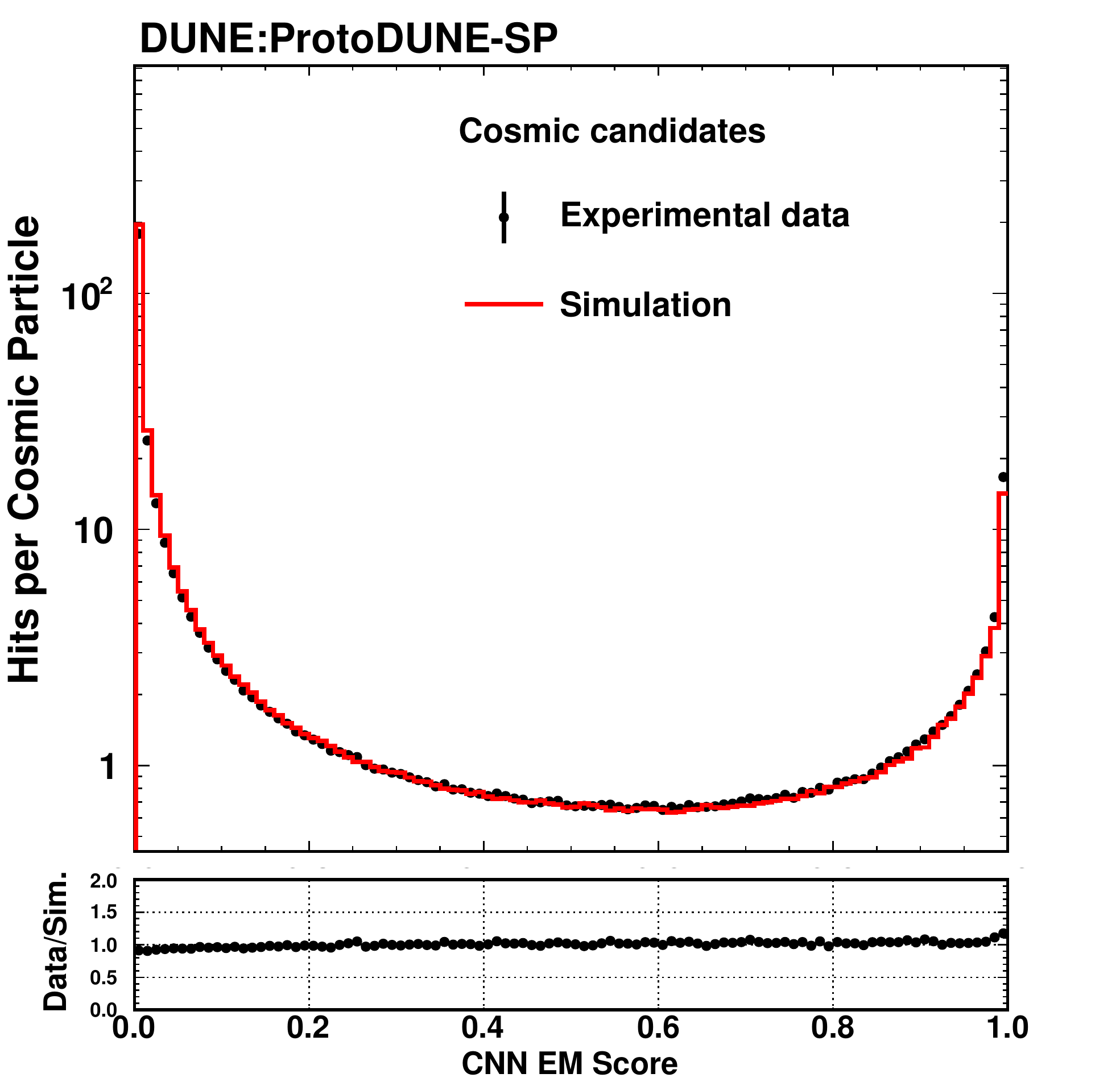}
    \caption{The CNN shower classifier scores for cosmic-ray muon hits from experimental data (black) and simulation (red). The error bars on the data are statistical.}
    \label{fig:cnn_cosmics}
\end{figure}

Firstly, the hit-level classification was studied. Figure~\ref{fig:cnn_cosmics} shows the CNN shower output score for cosmic-ray hits in experimental data (black) and simulation (red), and demonstrates the high level of agreement between the two samples. The peak in the score distribution close to one can be attributed to hits from the numerous delta-ray electrons produced by high energy muons, such as those shown previously in Fig.~\ref{fig:event_r5387}. The results of the hit-level classification, obtained by measuring the fraction of hits below a threshold of 0.72, are given in Table~\ref{tab:cosmicresults}.

\begin{figure}[htp!]
    \centering
    \includegraphics[width=0.48\textwidth]{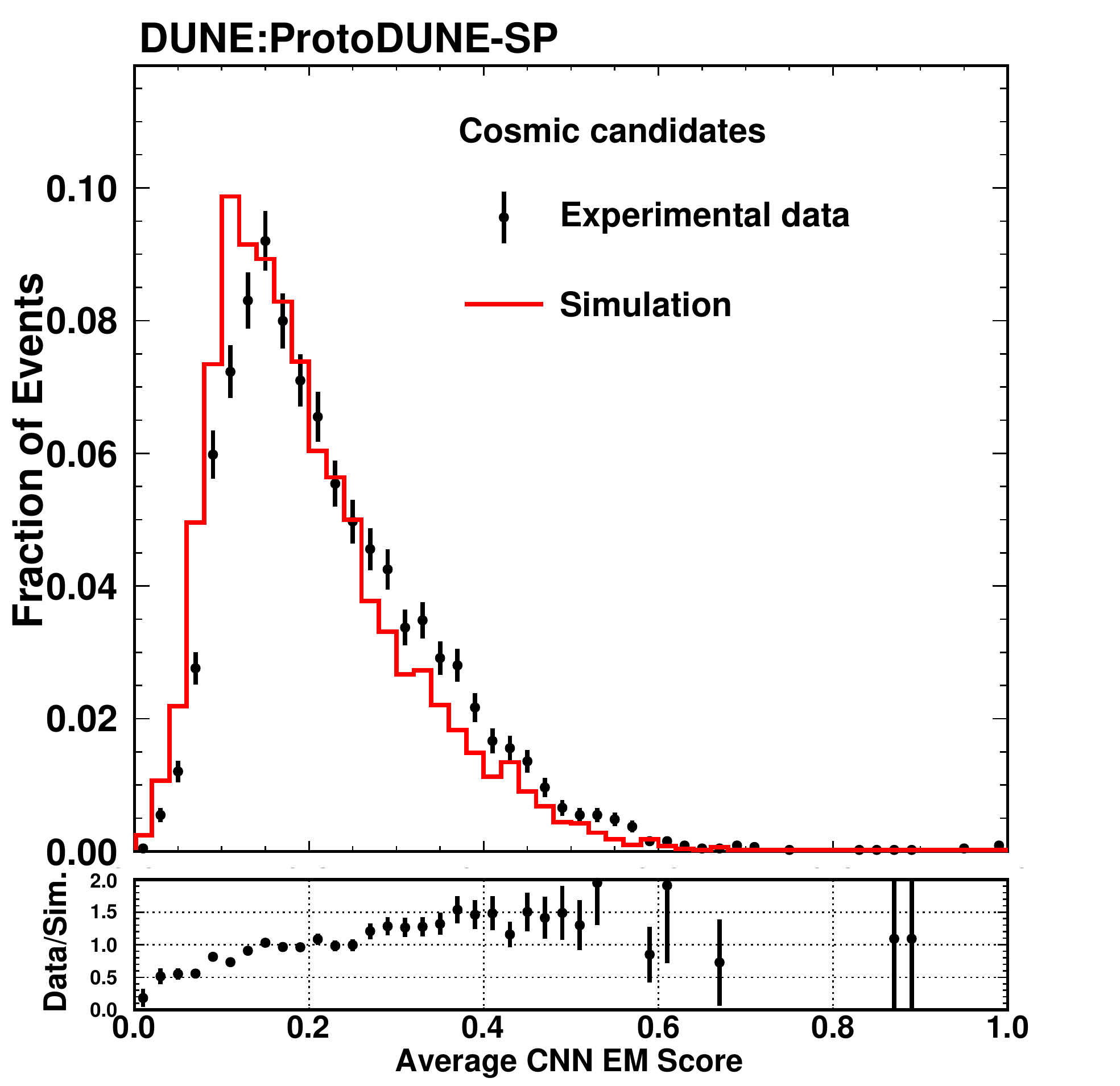}
    \caption{The average CNN shower classifier scores for cosmic-ray muons. The error bars on the experimental data are statistical.}
    \label{fig:avgcnn_cosmics}
\end{figure}

Figure~\ref{fig:avgcnn_cosmics} shows the particle-level comparison of the average CNN shower score for the cosmic-ray muons in experimental data and simulation. As expected, the distributions are peaked close to zero, and the experimental data distribution is slightly shifted compared to the simulation. However, when applying the threshold of 0.81 to classify the cosmic rays as track-like, both the performance and agreement between experimental data and simulation is excellent, as shown in Table~\ref{tab:cosmicresults}. 

\begin{table}[htp!]
    \centering
    \caption{The fraction of correctly classified cosmic-ray muon hits and particles using the CNN measured for experimental data and simulation. The errors represent the statistical uncertainties calculated using the Clopper-Pearson method~\cite{clopperPearson}.}
    \begin{tabular}{c|c|c|c}
         \hline \rule{0pt}{3ex}
         \multirow{2}{*}{Stage} & \multicolumn{2}{c|}{Correctly classified (\%)} & \multirow{2}{*}{Data/Simulation} \\[2pt] 
         & \hspace{0.4cm}Data\hspace{0.4cm} & Simulation \\[2pt] \hline
         \rule{0pt}{3ex}Hits & 85.6$\pm0.0$ & 87.3$\pm0.0$ & 0.980$\pm 0.000$ \\
         \rule{0pt}{3ex}Particles & 99.8$\pm0.1$ & 100.0$^{+0.0}_{-0.1}$ & 0.998$\pm0.002$ \\[4pt] 
         \hline
    \end{tabular}
    \label{tab:cosmicresults}
\end{table}

\subsection{Charged particle test beam}\label{sec:beam_results}
In the case of particles originating from the charged particle beam in the experimental data samples, the beam instrumentation~\cite{Abi:2020mwi} can be used to provide an effective truth source to which the results from the CNN can be compared. For simulation samples we use the truth information to get the primary beam particle species information. This allows the shower score distributions from the CNN to be compared between experimental data and simulation for different particle species. The reconstructed particles with angles inconsistent with the beam direction and that arrive out-of-time with the beam can be assumed to be cosmic muons. Note that at 1\,GeV/$c$ beam momentum, $\pi^{+}$ and $\mu^{+}$ are indistinguishable using the beam instrumentation information. A 1\,GeV/$c$ $\mu^{+}$ is expected to stop in the middle of the detector around z = 380\,cm, where the z axis is horizontal. A 1\,GeV/$c$ $\pi^{+}$ will most likely interact with the argon nucleus before stopping because of the relatively short interaction length ($\sim$100\,cm). We identify an event as a pion if the reconstructed track end z position is less than 100\,cm and as a muon if the end z position is greater than 300\,cm for events identified by the beam instrumentation as either pions or muons. We require the number of collection plane hits in the reconstructed shower should be greater than 200 for the positron candidate events in order to remove events with an incompletely reconstructed shower. This cut is not applied to the other three particle species. Table~\ref{tab:events} shows the numbers of events after the beam quality and number of hits cuts for beam pions, muons, protons and positrons and the purity of the selected samples based on the truth information in the simulation. 
\begin{table}[!htp]
  \center
  \caption{Numbers of events after the beam quality and number of hits selection criteria shown for experimental data and simulation.}
  \label{tab:events}
  \begin{tabular}{c|c|c|c|c}
    \hline
     & $\pi^{+}$ & $\mu^{+}$ & $p$ & $e^{+}$ \\
    \hline
    Data & 5402 & 1228 & 9364 & 9106 \\
    Simulation & 16612 & 1305 & 23660 & 42245 \\
    Simulation purity & 84.4\% & 86.7\% & 99.8\% & 97.7\%\\
    \hline
  \end{tabular}
\end{table}

\begin{figure*}[htp!]
    \centering
    \begin{subfigure}[b]{0.48\textwidth}
        \centering
        \includegraphics[width=\textwidth]{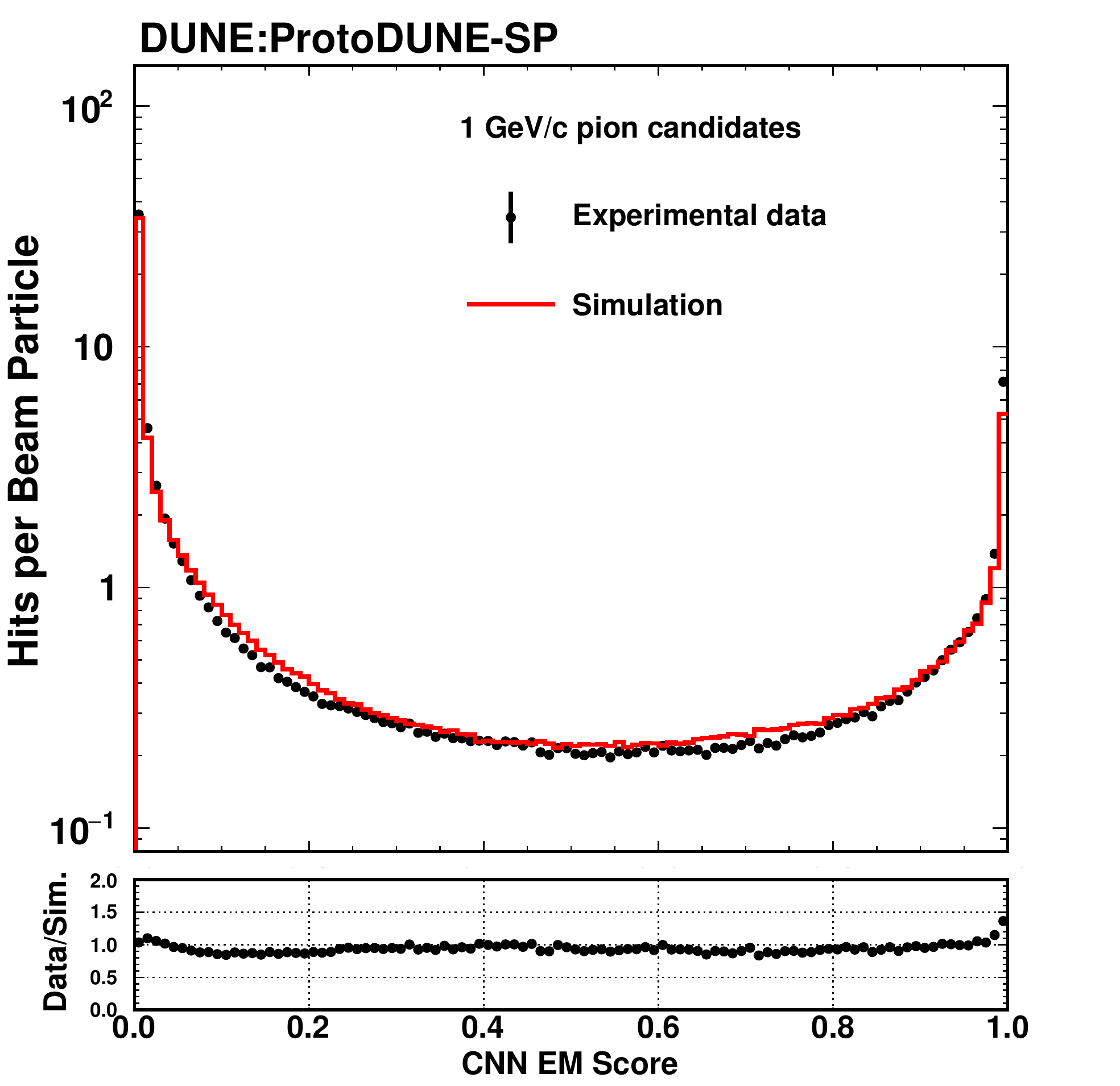}
        \caption{pion}
        \label{fig:pion}
    \end{subfigure}
    \begin{subfigure}[b]{0.48\textwidth}
        \centering
        \includegraphics[width=\textwidth]{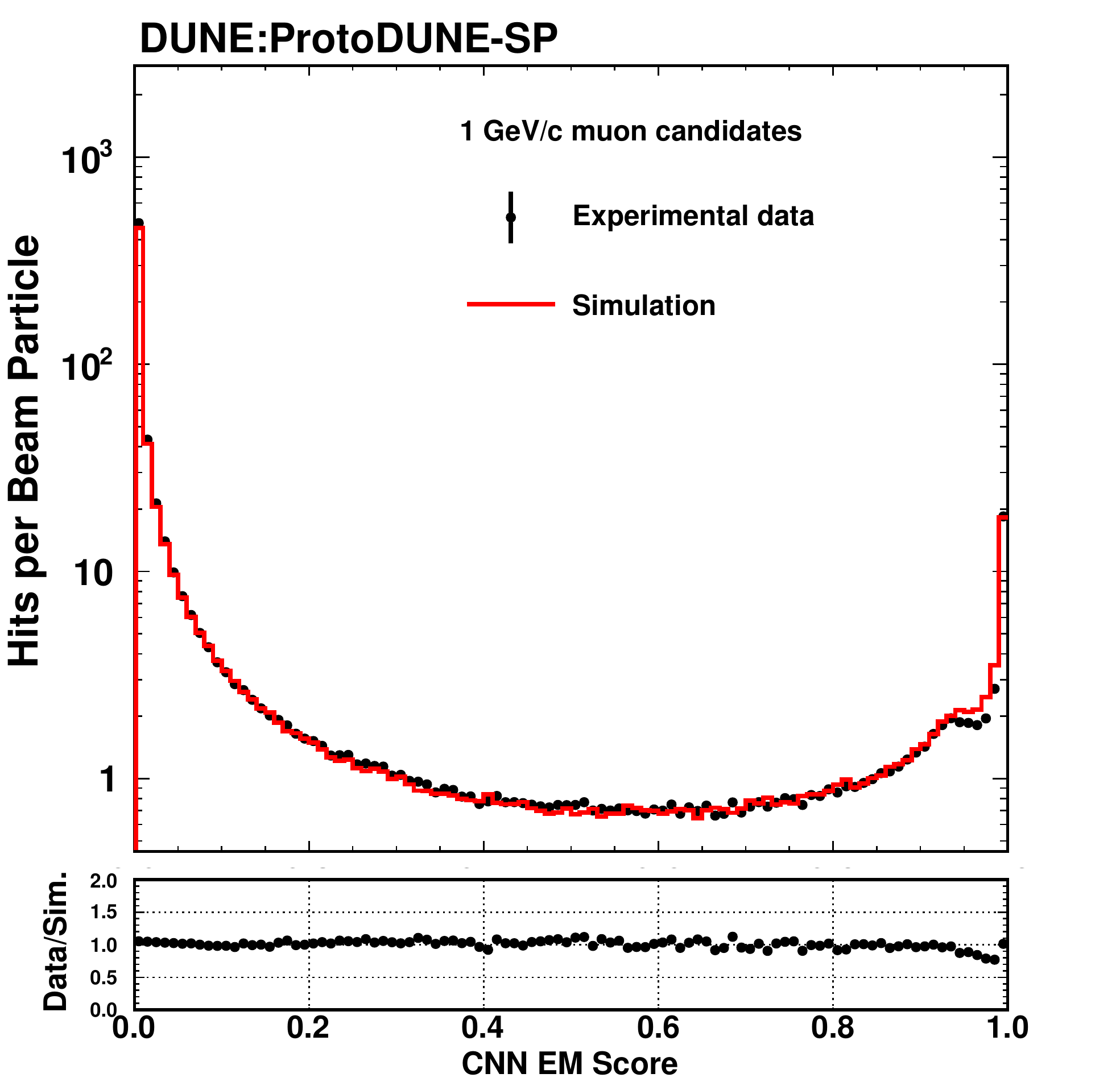}
        \caption{muon}
        \label{fig:muon}
    \end{subfigure}
    \begin{subfigure}[b]{0.48\textwidth}
        \centering
        \includegraphics[width=\textwidth]{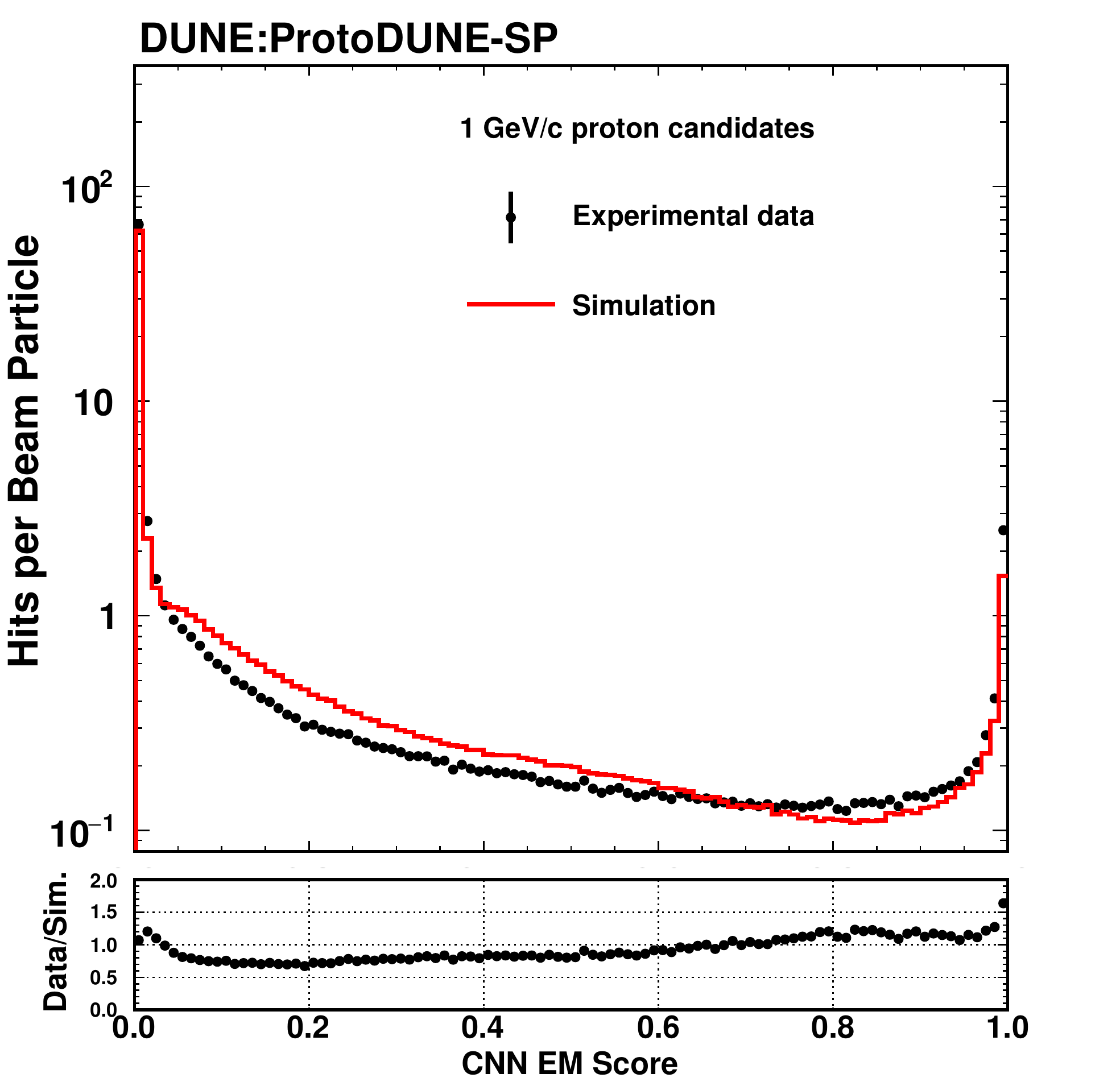}
        \caption{proton}
        \label{fig:proton}
    \end{subfigure}
    \begin{subfigure}[b]{0.48\textwidth}
        \centering
        \includegraphics[width=\textwidth]{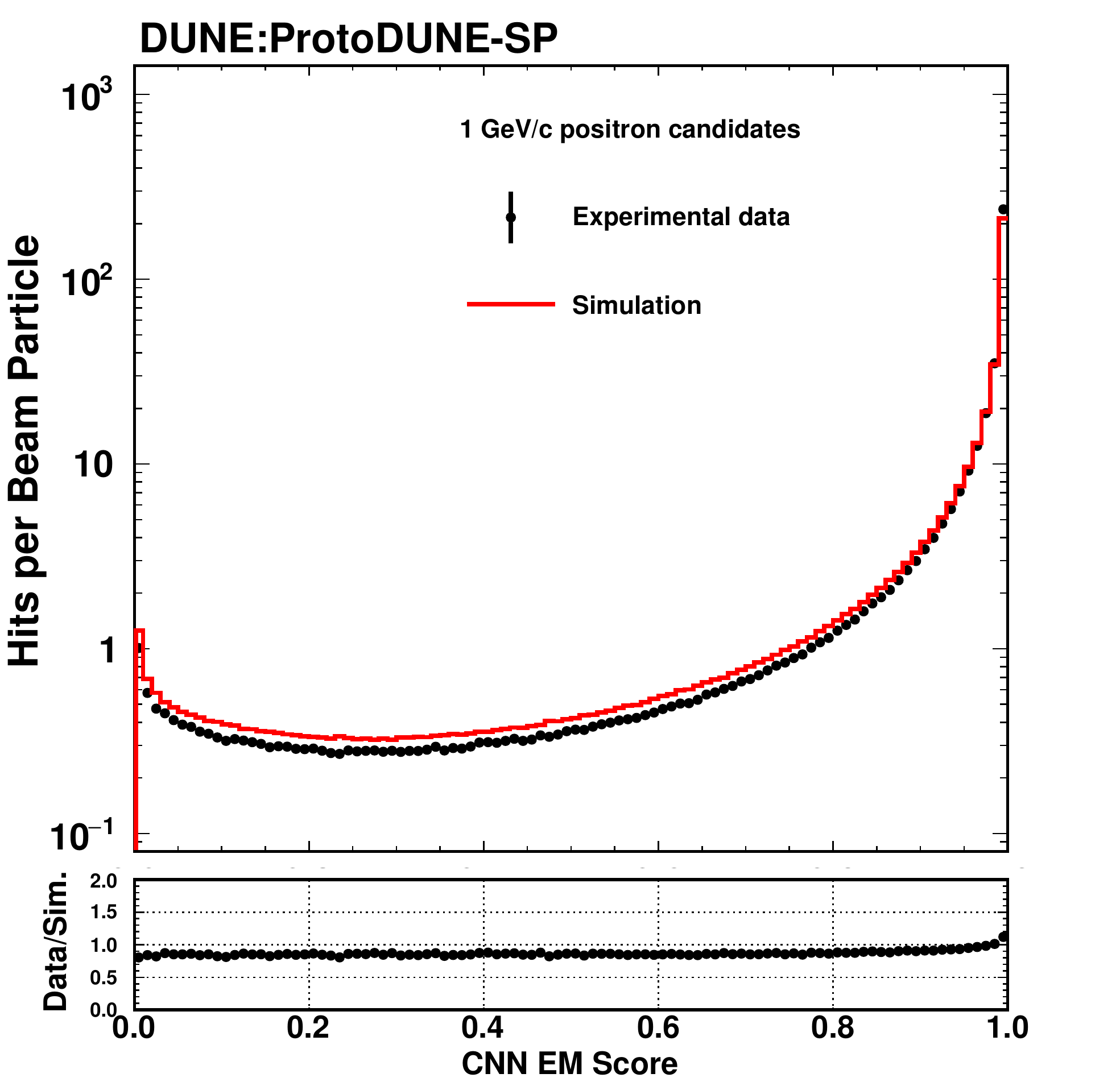}
        \caption{positron}
        \label{fig:positron}
    \end{subfigure}
    \caption{Shower classifier scores for different particle species in the ProtoDUNE-SP beam. The error bars on the experimental data are statistical.}
    \label{fig:cnn_beam}
\end{figure*}

Figure~\ref{fig:cnn_beam} shows the distribution of shower classifier score for each individual hit in the beam pions, muons, protons, and positrons. The data in all of the beam particle distributions are normalised by the number of triggered beam particles of the given flavour after the beam quality and number of hits cuts. There is a reasonable agreement between the experimental data and simulation in terms of the shower score distributions for each particle species. To quantify the efficiency to select track--like and shower--like hits, Table~\ref{tab:frac_selected} lists the fraction of individual hits selected into the appropriate category for each sample in experimental data and simulation for a selection threshold of 0.72. The difference between the selected fraction in each case is an estimate of the systematic uncertainty associated with hit-by-hit selection. The class used for the selection in each sample is also given in Table~\ref{tab:frac_selected}. The fractional difference between experimental data and simulation varies based on the particles species, and falls in the range of 1-2\%.

\begin{table*}[htp!]
	\centering
	\caption{Fraction of hits classified into appropriate class for different samples in ProtoDUNE-SP data  and simulation. The statistical uncertainties on the fractions and ratios are negligible.}
	\bgroup 
	\begin{tabular}{c|c|c|c|c}
                \hline
		Hit Source  & Class  & Data Fraction (\%) & Simulation Fraction (\%) & Data / Simulation\\\hline
		Pion        & Track  & 78.7             & 80.3               & 0.98     \\
		Muon        & Track  & 92.7             & 92.0               & 1.01     \\
		Proton      & Track  & 93.0             & 94.5               & 0.98     \\
		Positron    & Shower & 93.0             & 91.4               & 1.02     \\
                \hline
	\end{tabular}
	\egroup
	\label{tab:frac_selected}
\end{table*}

\begin{figure*}[htp!]
    \centering
    \begin{subfigure}[b]{0.48\textwidth}
        \centering
        \includegraphics[width=\textwidth]{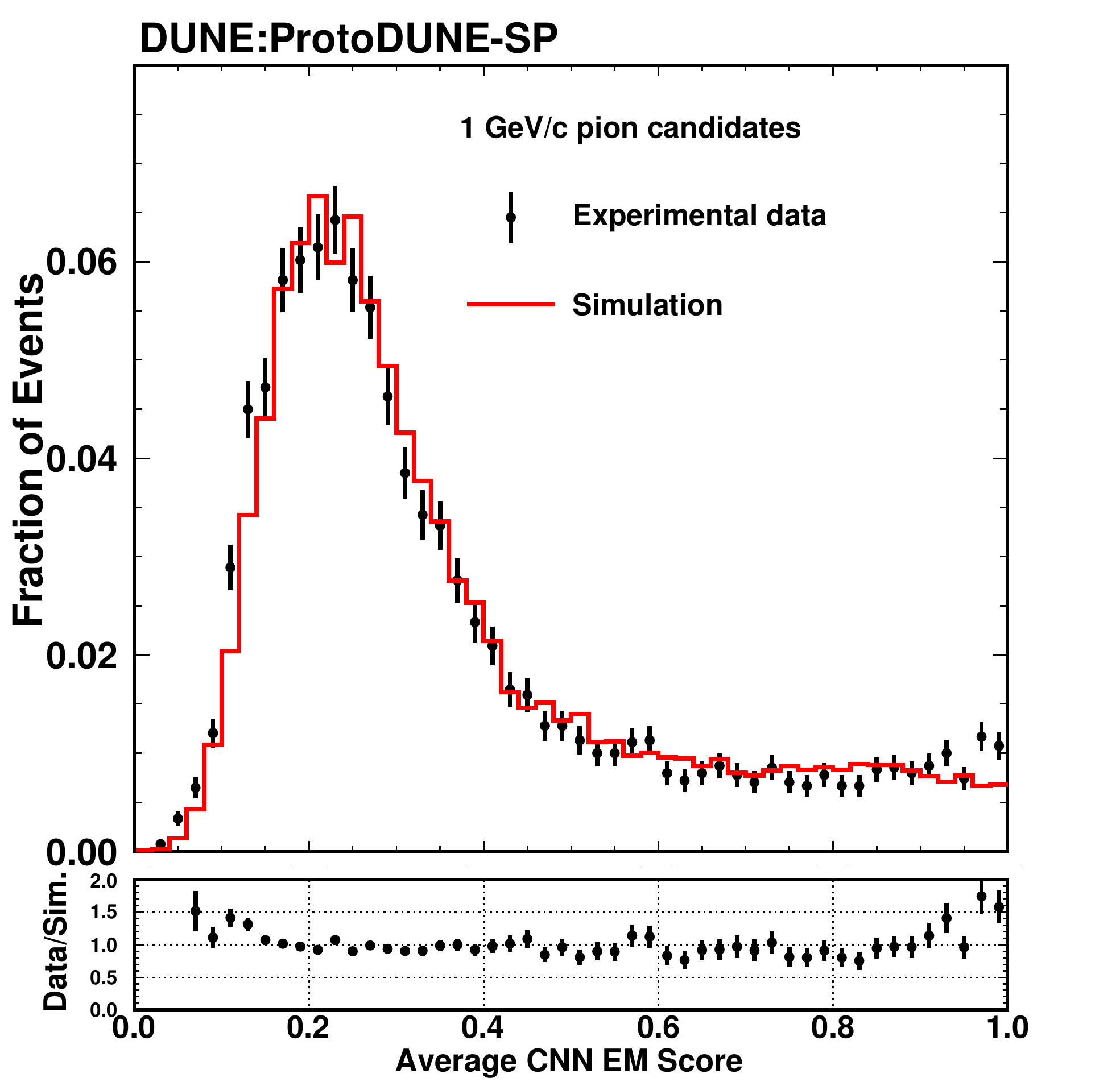}
        \caption{pion}
        \label{fig:pionpart}
    \end{subfigure}
    \begin{subfigure}[b]{0.48\textwidth}
        \centering
        \includegraphics[width=\textwidth]{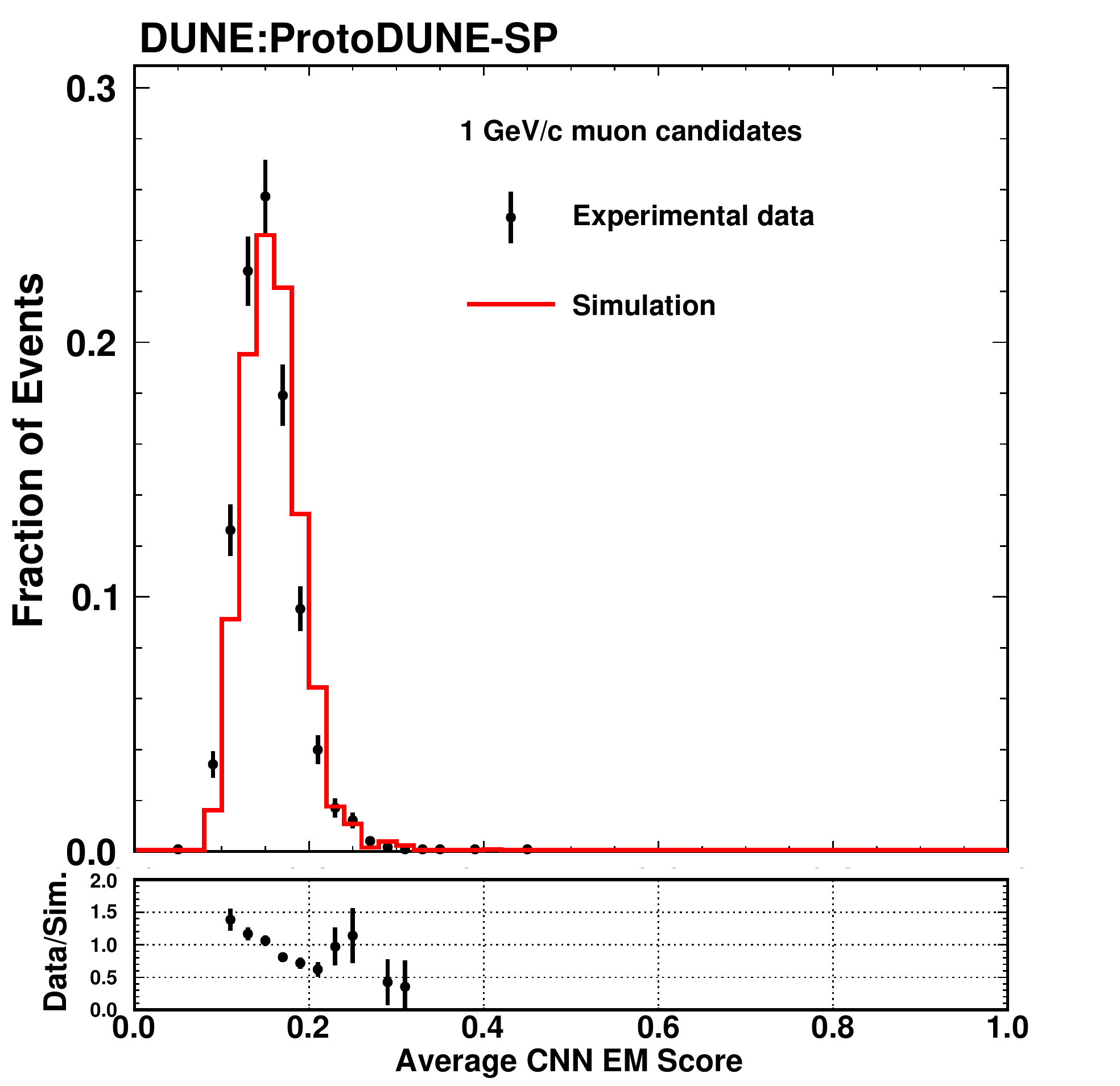}
        \caption{muon}
        \label{fig:muonpart}
    \end{subfigure}
    \begin{subfigure}[b]{0.48\textwidth}
        \centering
        \includegraphics[width=\textwidth]{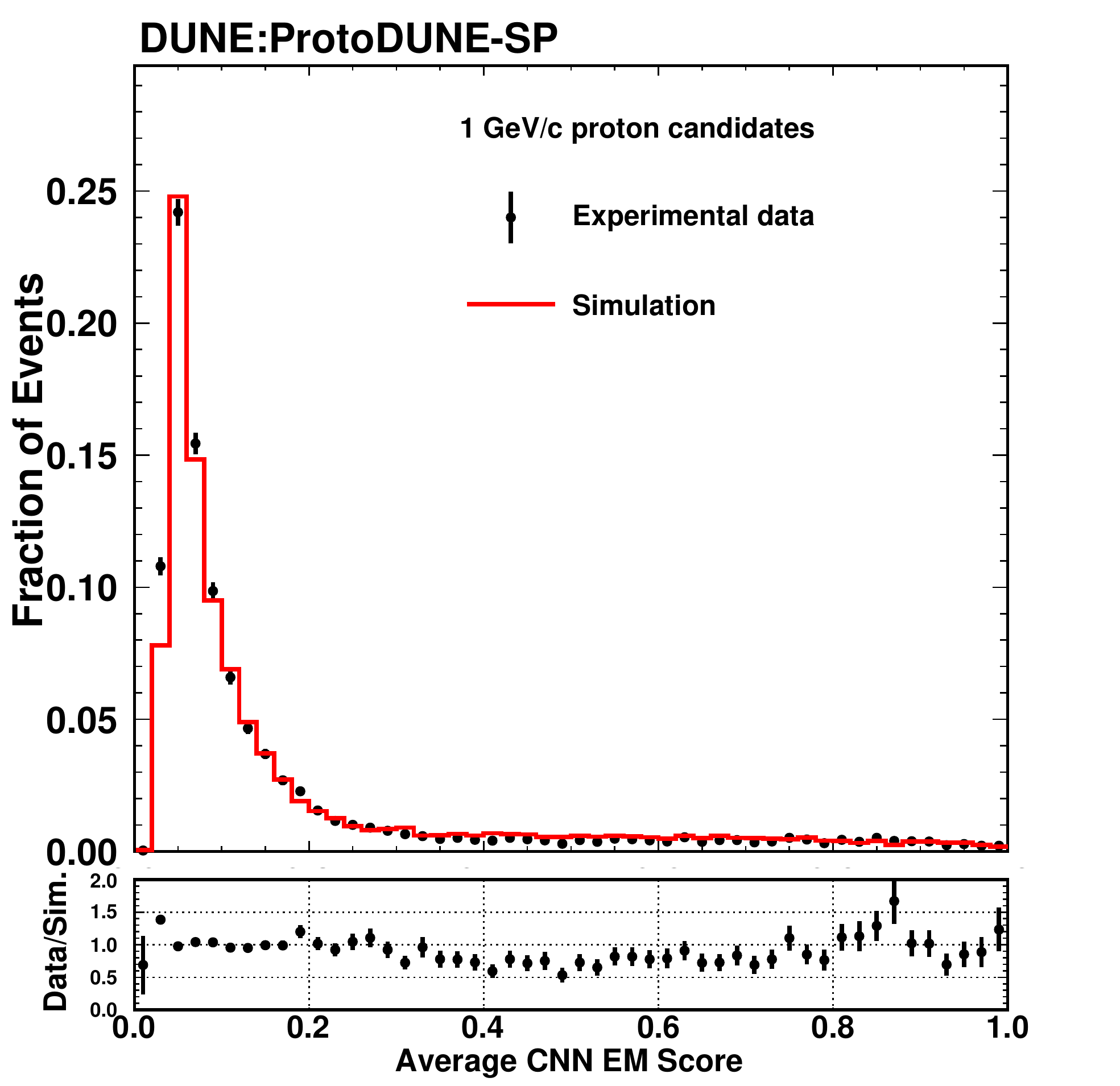}
        \caption{proton}
        \label{fig:protonpart}
    \end{subfigure}
    \begin{subfigure}[b]{0.48\textwidth}
        \centering
        \includegraphics[width=\textwidth]{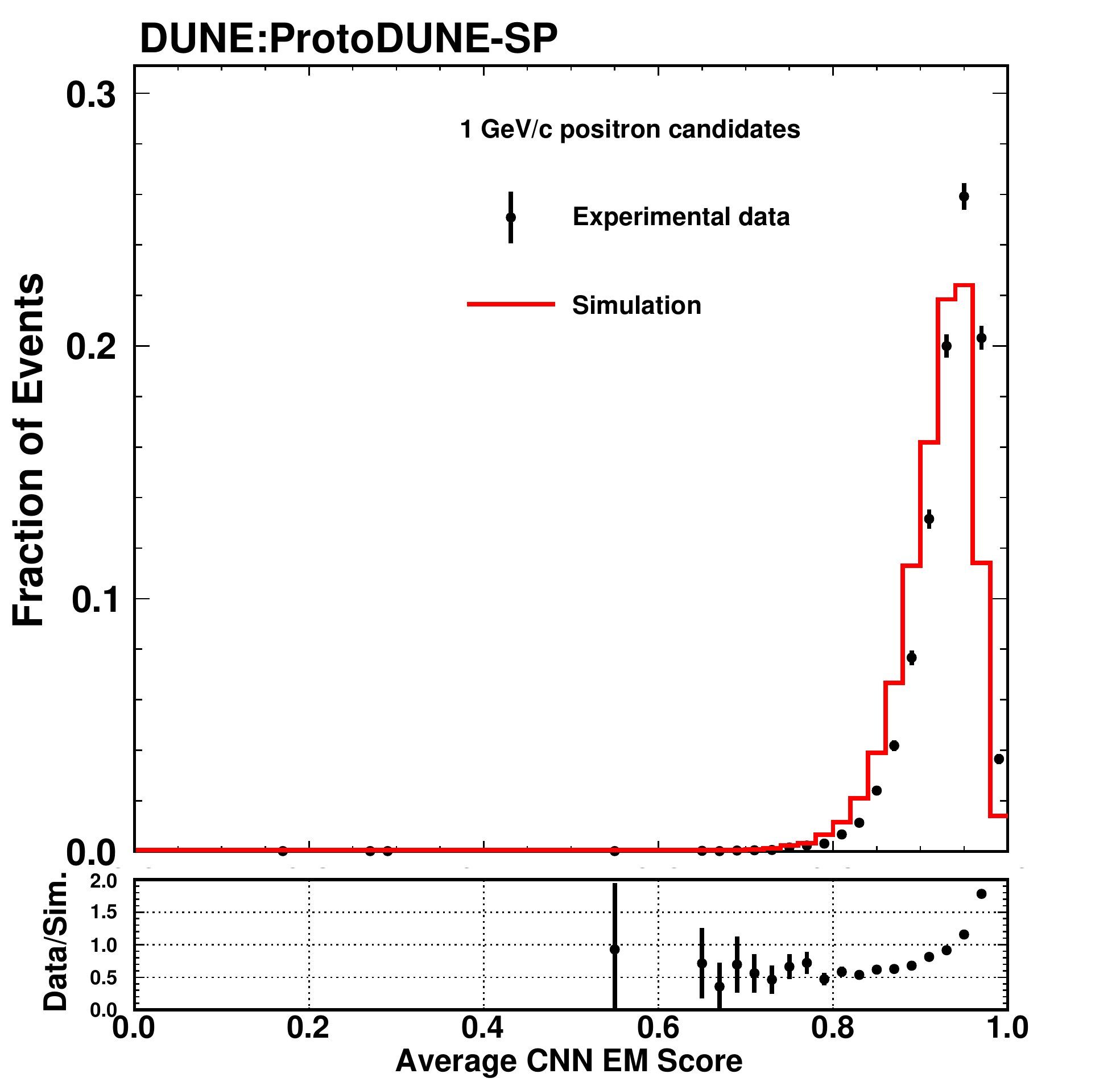}
        \caption{positron}
        \label{fig:positronpart}
    \end{subfigure}
    \caption{Average shower classifier scores for different particle species in the ProtoDUNE-SP beam. The error bars on the experimental data are statistical.}
    \label{fig:cnn_beampart}
\end{figure*}

Figure~\ref{fig:cnn_beampart} shows the distribution of the average shower classifier scores over all the hits in the reconstructed pion, proton, and positron particles. This average shower classifier score is what analysers normally use to identify a reconstructed particle as a track--like or shower--like particle. The distributions in each category are normalised to unit area. The experimental data and simulation distributions are in a reasonable agreement. There is a long tail in the average shower classifier score distribution for both the beam pions and protons. This tail is caused by the spatial distortion introduced by the SCE and is largely suppressed if we make the distributions using simulation sample without simulating SCE. There is a shift in the average shower classifier score for beam positrons between experimental data and simulation. There are slightly more hits in experimental data than in simulation for reconstructed positron events, making the experimental data hits more shower--like. It can be seen that the score distribution for the beam muons is more strongly peaked towards low scores than for cosmic-ray muons, shown in Fig.~\ref{fig:avgcnn_cosmics}, because they are significantly lower in energy and hence produce fewer delta rays. Table~\ref{tab:frac_selected_particles} lists the fraction of reconstructed particles selected into the appropriate category for each sample in experimental data and simulation for a selection threshold of 0.81. The fractional difference between experimental data and simulation is within 1\% for all particles species.

\begin{table*}[htp!]
	\centering
	\caption{Fraction of reconstructed particles classified into appropriate class for different samples in ProtoDUNE-SP data  and simulation. The errors represent the statistical uncertainties calculated using the Clopper-Pearson method~\cite{clopperPearson}.}
	\bgroup 
	\begin{tabular}{c|c|c|c|c}
                \hline
		Hit Source  & Class  & Data Fraction (\%) & Simulation Fraction (\%) & Data / Simulation\\\hline
		Pion        & Track  & 91.7$\pm$0.4       & 92.5$\pm$0.2     & 0.991$\pm$0.005    \\
		Muon        & Track  & 100$^{+0.0}_{-0.1}$  & 100$^{+0.0}_{-0.1}$ & 1.000$^{+0.000}_{-0.002}$    \\
		Proton      & Track  & 96.9$\pm$0.2       & 97.1$\pm$0.1     & 0.998$\pm$0.002    \\
		Positron   & Shower & 98.8$\pm$0.1       & 97.9$\pm$0.1     & 1.010$\pm$0.001    \\
                \hline
	\end{tabular}
	\egroup
	\label{tab:frac_selected_particles}
\end{table*}

\subsection{Michel electrons}
\begin{figure*}[htp!]
    \centering
    \begin{subfigure}[b]{0.48\textwidth}
        \centering
        \includegraphics[width=\textwidth]{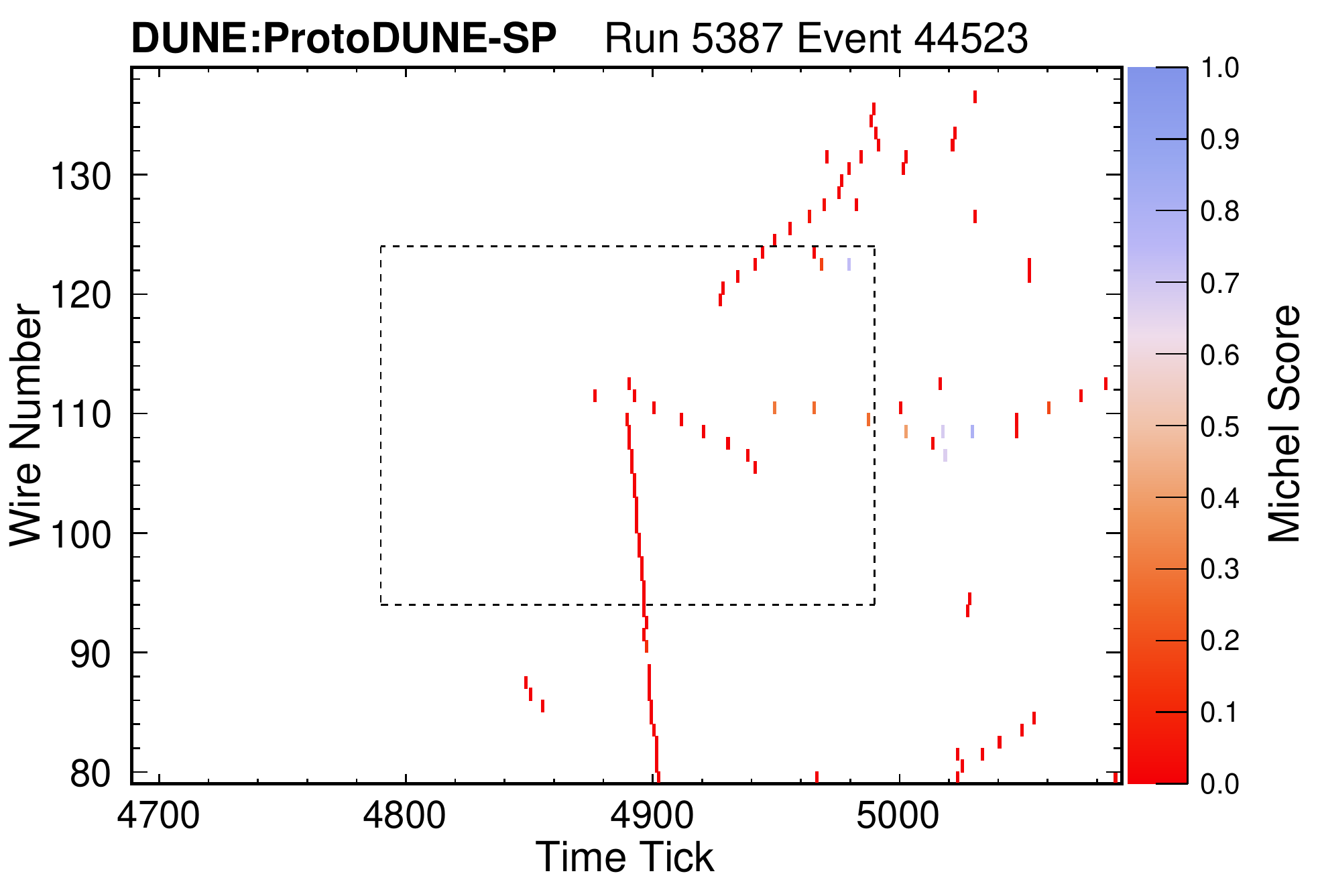}
        \caption{A pion candidate}
        \label{fig:michelpi}
    \end{subfigure}
    \begin{subfigure}[b]{0.48\textwidth}
        \centering
        \includegraphics[width=\textwidth]{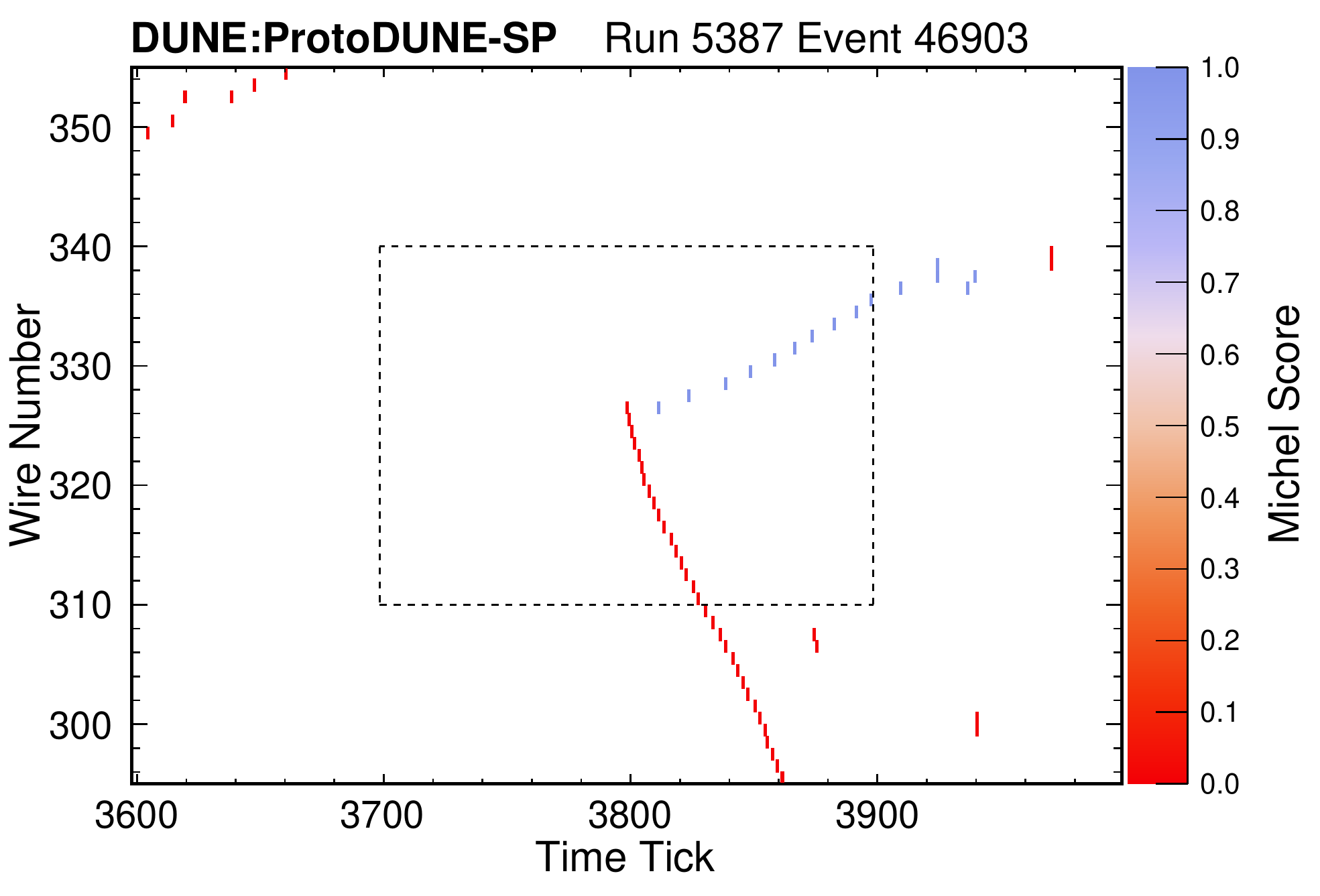}
        \caption{A muon candidate}
        \label{fig:michelmu}
    \end{subfigure}
    \caption{CNN Michel score for reconstructed primary beam particles and secondary particles in a reconstructed pion (left) and muon (right) particle. Each coloured pixel shows the peak time and wire number of a hit. The box surrounding the track end point is used to select the daughter hits. The average daughter Michel score is 0.005 for the pion event and 1.000 for the muon event.}
    \label{fig:michelscore}
\end{figure*}

\begin{figure*}[htp!]
    \centering
    \begin{subfigure}[b]{0.48\textwidth}
        \centering
        \includegraphics[width=\textwidth]{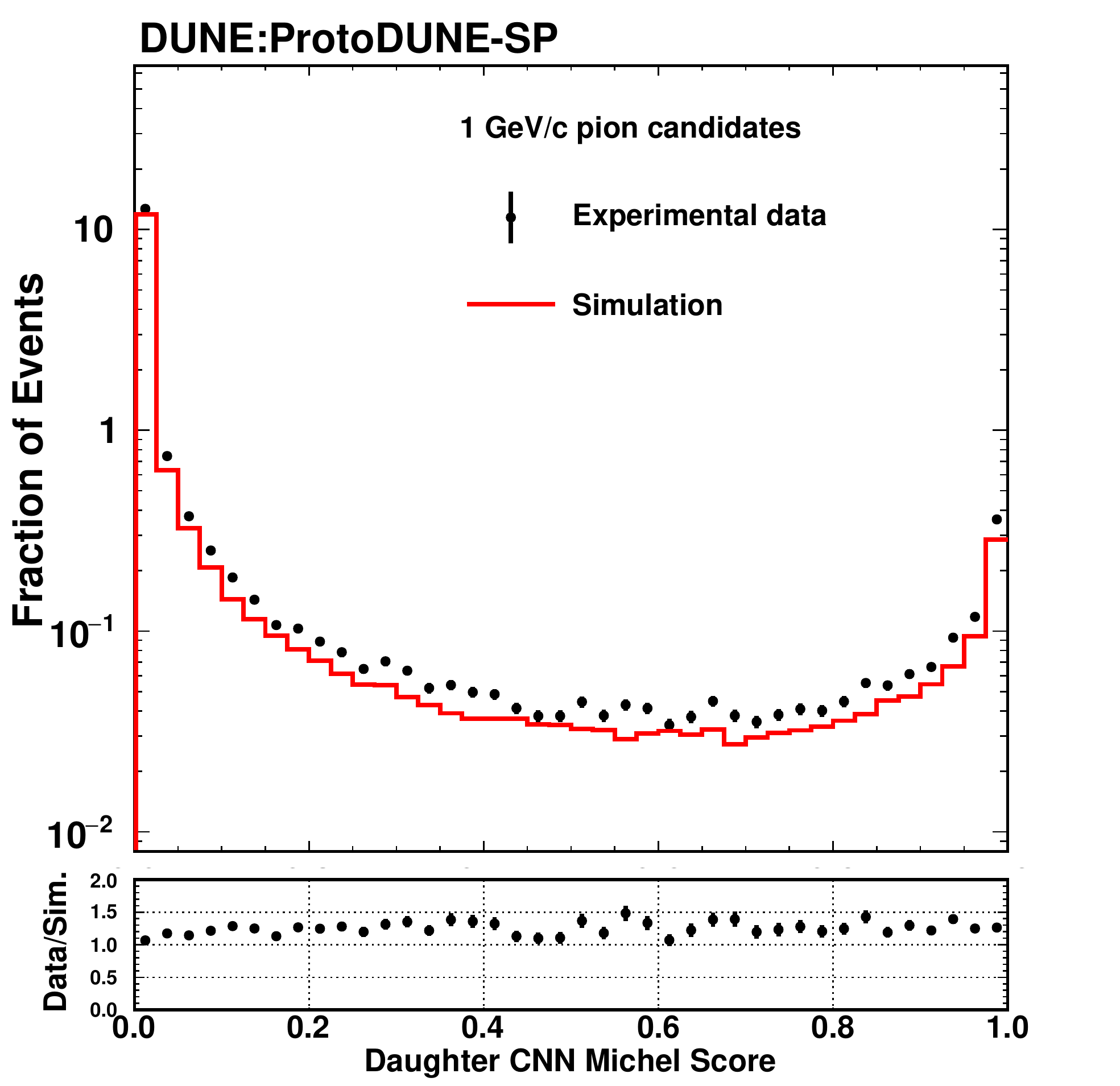}
        \caption{Pion candidates}
        \label{fig:hitmichelpi2}
    \end{subfigure}
    \begin{subfigure}[b]{0.48\textwidth}
        \centering
        \includegraphics[width=\textwidth]{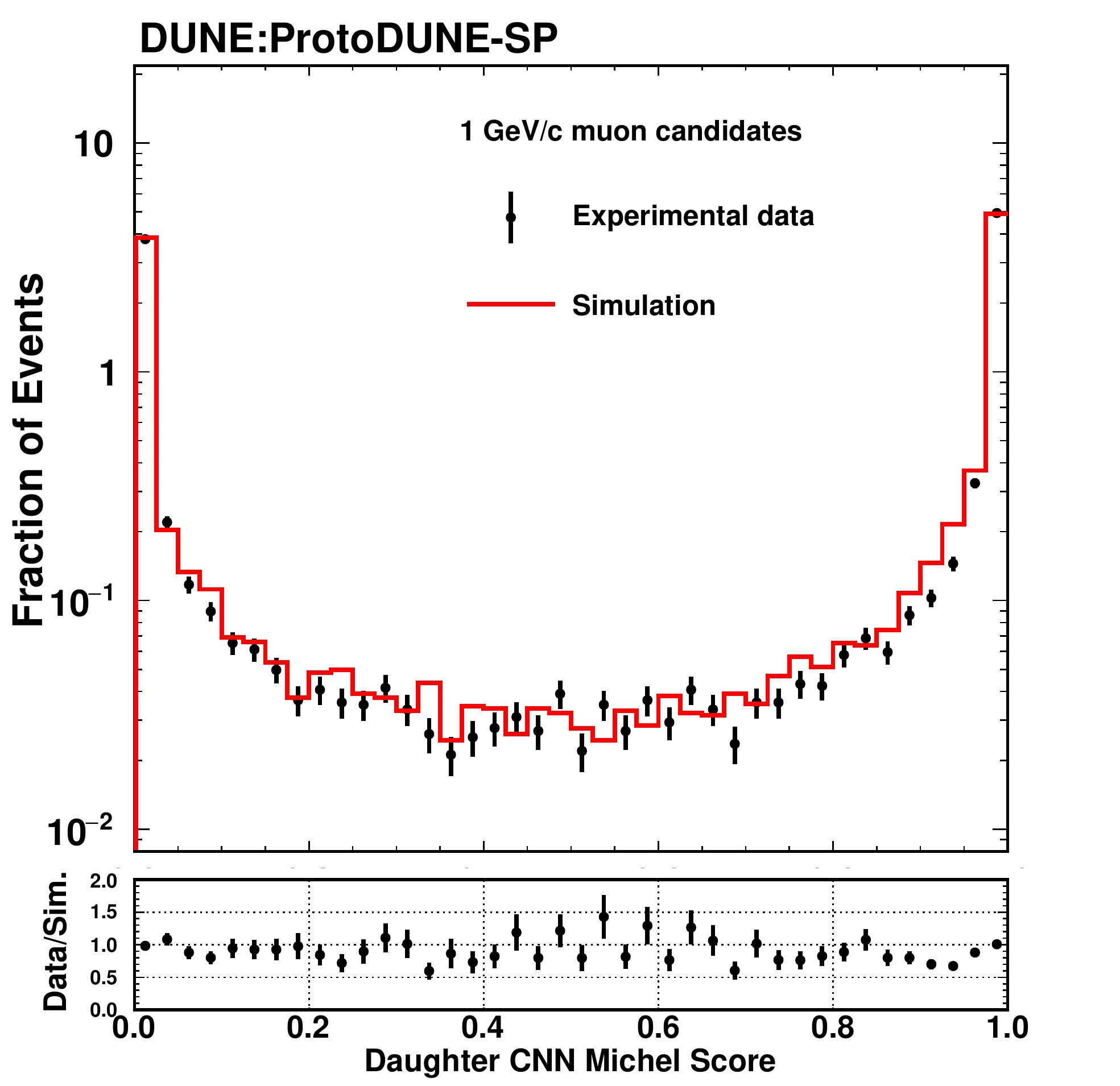}
        \caption{Muon candidates}
        \label{fig:hitmichelmu2}
    \end{subfigure}
    \caption{CNN Michel score for the daughter hits in the 30 wires $\times$ 200 ticks window centred around the reconstructed track end point of the pion (left) and muon (right) particles.}
    \label{fig:hitmichelscore2}
\end{figure*}

\begin{figure*}[htp!]
    \centering
    \begin{subfigure}[b]{0.48\textwidth}
        \centering
        \includegraphics[width=\textwidth]{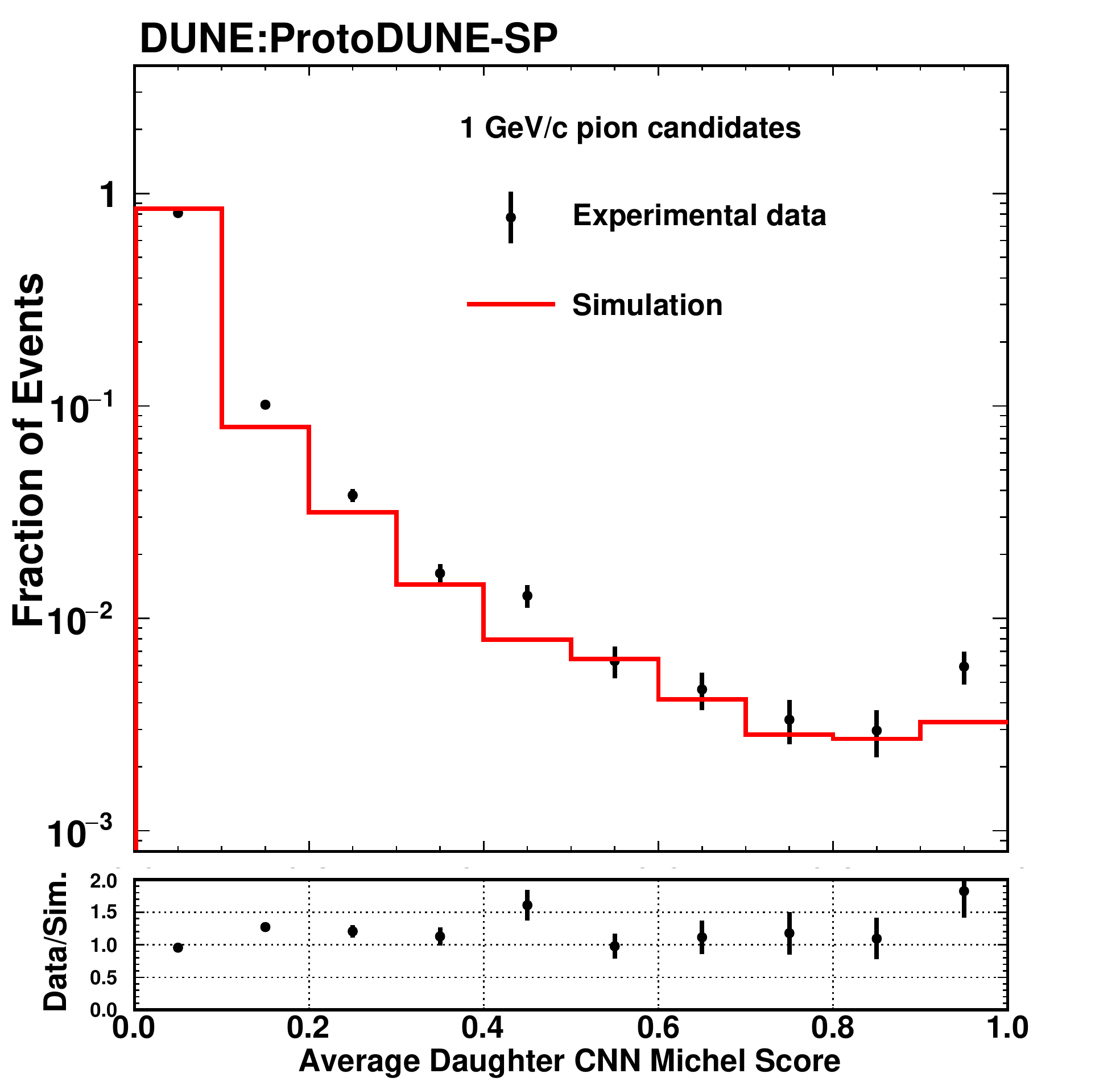}
        \caption{Pion candidates}
        \label{fig:michelpi2}
    \end{subfigure}
    \begin{subfigure}[b]{0.48\textwidth}
        \centering
        \includegraphics[width=\textwidth]{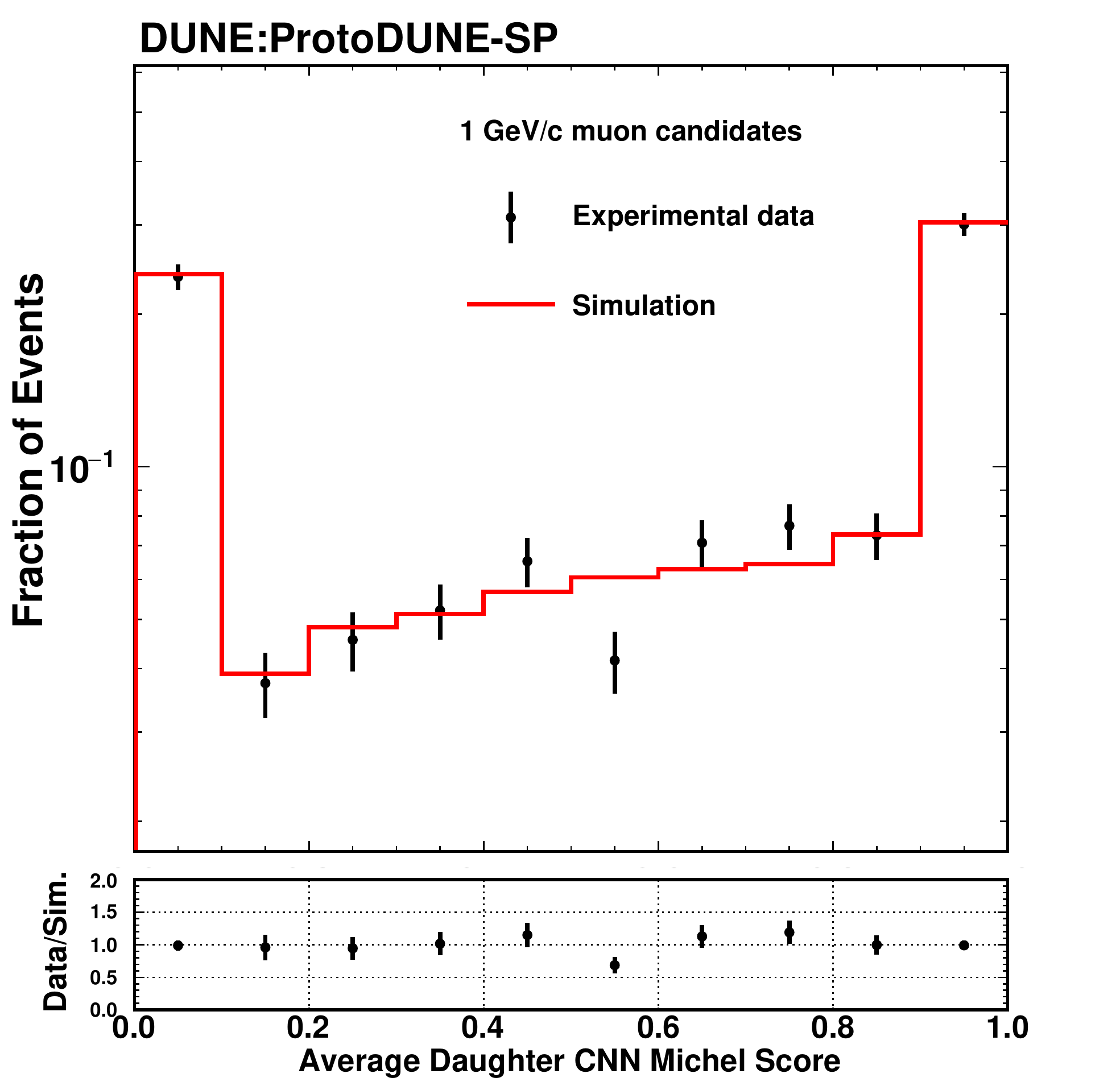}
        \caption{Muon candidates}
        \label{fig:michelmu2}
    \end{subfigure}
    \caption{Average CNN Michel score over the daughter hits in the 30 wires $\times$ 200 ticks window centred around the reconstructed track end point of the pion (left) and muon (right) particles.}
    \label{fig:michelscore2}
\end{figure*}

To validate the performance of the CNN Michel score calculation, we examine the Michel score of hits around the muon and pion track end point. Hits around the muon end points are most likely from the Michel electron which are expected to have a high Michel score. We define a window of 30 wires $\times$ 200 ticks (approximately $15\times 16$ cm$^{2}$) centred around the reconstructed track end point projected on the collection plane to select daughter hits. Hits from the secondary particles produced by the pion interaction are expected to have a low Michel score as shown in Fig.~\ref{fig:michelpi}. The Michel hits from the muon decay are expected to have a high Michel score as shown in Fig.~\ref{fig:michelmu}. Hits on the primary beam track or on another track that is longer than 25 cm are excluded to remove the contributions from primary beam particle and cosmic ray muons. Figures~\ref{fig:hitmichelscore2} and ~\ref{fig:michelscore2} show the hit-level and particle level comparison of the CNN Michel score over daughter hits in the reconstructed pion and muon particles. 

The results of the hit-level and event-level classification, obtained using a threshold of 0.19, are given in Tables~\ref{tab:frac_selected_particles_michelhit} and ~\ref{tab:frac_selected_particles_michel}, respectively. The threshold is chosen to maximise the product of selection efficiencies of pions and muons. We are able to select 73\% of the $\mu^{+}$ events while rejecting 90--92\% of the $\pi^{+}$ events using the average Michel score. The fractional difference between experimental data and simulation falls in the range of 1--2\%. Efficient identification of Michel electrons provides crucial information on particle identification and kinematic reconstruction. It allows the separation between $\mu^{+}$ and $\mu^{-}$ because 70\% of the $\mu^{-}$s are captured while most of the $\mu^{+}$s decay into Michel electrons. It also allows the identification of stopping $\pi^{+}$ which goes through the decay chain $\pi^{+}\rightarrow\mu^{+}\rightarrow e^{+}$. The momentum of those stopping pions can be reconstructed either through track range or using calorimetric information, which can be used to reconstruct the full kinematics of the final state particles.
\begin{table*}[htp!]
	\centering
	\caption{Fraction of daughter hits classified into appropriate class for different samples in ProtoDUNE-SP data  and simulation. The statistical uncertainties on the fractions and ratios are negligible.}
	\bgroup 
        \begin{tabular}{c|c|c|c|c}
                \hline
		Hit Source     & Class      & \makecell{Data \\ Fraction} (\%) & \makecell{Simulation \\ Fraction} (\%) & Data / Simulation\\\hline
		Pion daughters & Non-Michel--like  & 87.6     & 89.2       & 0.982     \\
		Muon daughters & Michel--like  & 59.8     & 60.2       & 0.993     \\
                \hline
	\end{tabular}
	\egroup
	\label{tab:frac_selected_particles_michelhit}
\end{table*}

\begin{table*}[htp!]
	\centering
	\caption{Fraction of reconstructed particles classified into appropriate class for different samples in ProtoDUNE-SP data  and simulation. The errors represent the statistical uncertainties calculated using the Clopper-Pearson method~\cite{clopperPearson}.}
	\bgroup 
        \begin{tabular}{c|c|c|c|c}
                \hline
		Hit Source     & Class      & \makecell{Data \\ Fraction} (\%) & \makecell{Simulation \\ Fraction} (\%) & Data / Simulation\\\hline
		Pion daughters & Non-Michel--like  & 90.4$\pm$0.4     & 92.2$\pm$0.2       & 0.980$\pm$0.005     \\
		Muon daughters & Michel--like  & 73.2$\pm$1.3     & 72.6$\pm$1.3       & 1.009$^{+0.025}_{-0.026}$     \\
                \hline
	\end{tabular}
	\egroup
	\label{tab:frac_selected_particles_michel}
\end{table*}

\section{Conclusion}
In this paper, we described an effective hit tagging algorithm for track, shower, and Michel electron hit classification based on a convolutional neural network, using a small patch approach. 
This algorithm is shown to give good agreement in selection efficiencies, of around 1--2\%, between experimental data and simulation for cosmic rays and 1\,GeV/$c$ test-beam interactions for a hit-by-hit event selection. When combined with the full event reconstruction (which includes a BDT-based classifier) and applied to the hits of each reconstructed particle, the CNN refines the track and shower classification to produce highly efficient selections that agree within 1\% between experimental data and simulation. Additionally, this network also provides a method to select Michel electrons, which helps with the particle identification and kinematic reconstruction. This algorithm is being widely used within ongoing ProtoDUNE-SP data analyses, including pion cross-section analyses and detector calibrations. 

\section*{Acknowledgements}




%
%
The ProtoDUNE-SP detector was constructed and operated on the CERN Neutrino Platform.
We gratefully acknowledge the support of the CERN management, and the
CERN EP, BE, TE, EN and IT Departments for NP04/Proto\-DUNE-SP.
%
%
This document was prepared by the DUNE collaboration using the
resources of the Fermi National Accelerator Laboratory 
(Fermilab), a U.S. Department of Energy, Office of Science, 
HEP User Facility. Fermilab is managed by Fermi Research Alliance, 
LLC (FRA), acting under Contract No. DE-AC02-07CH11359.
%
%
This work was supported by
CNPq,
FAPERJ,
FAPEG and 
FAPESP,                         Brazil;
CFI, 
IPP and 
NSERC,                          Canada;
CERN;
M\v{S}MT,                       Czech Republic;
ERDF, 
H2020-EU and 
MSCA,                           European Union;
CNRS/IN2P3 and
CEA,                            France;
INFN,                           Italy;
FCT,                            Portugal;
NRF,                            South Korea;
CAM, 
Fundaci\'{o}n ``La Caixa'',
Junta de Andaluc\'ia-FEDER,
MICINN, and
Xunta de Galicia,               Spain;
SERI and 
SNSF,                           Switzerland;
T\"UB\.ITAK,                    Turkey;
The Royal Society and 
UKRI/STFC,                      United Kingdom;
DOE and 
NSF,                            United States of America.
%
%
This research used resources of the 
National Energy Research Scientific Computing Center (NERSC), 
a U.S. Department of Energy Office of Science User Facility 
operated under Contract No. DE-AC02-05CH11231.
%

\bibliographystyle{unsrt}
\bibliography{sectionbib}

\end{document}